\documentclass[aps,preprint]{revtex4}

\usepackage[dvips]{graphicx}

\newcommand{\ch}{{\cal H}}
\newcommand{\ce}{{\cal E}}

\begin{document}
\date{\today}
\title{
Width bifurcation and dynamical phase transitions \\ in open quantum systems
}

\author{Hichem Eleuch$^{1,2,*}$ and Ingrid Rotter$^{3,**}$}

\address{$^1$Ecole Polytechnique, C.P. 6079, succ. Centre ville, 
Montr\'eal (QC), Canada H3C 3A7}

\address{$^2$Universit\'e de Montr\'eal, C.P. 6128, Succ. Centre-Ville,
Montr\'eal (QC), H3C 3J7 Canada}

\address{$^3$Max Planck Institute for the Physics of Complex
Systems, D-01187 Dresden, Germany }

\begin{abstract}

The states of an open quantum system are coupled via the environment
of scattering wavefunctions. The complex coupling coefficients $\omega$
between system and environment arise from the
principal value integral and the residuum. At high level
density where the resonance states overlap, the dynamics of the system
is determined by exceptional points. At these points, the eigenvalues
of two states are equal and the corresponding eigenfunctions are linearly
dependent. It is shown in the present paper that Im$(\omega)$ and 
Re$(\omega)$ influence the system properties differently in the
surrounding of exceptional points.  Controlling the system by a parameter, 
the eigenvalues avoid crossing in energy near an exceptional point 
under the  influence of Re$(\omega)$
in a similar manner as it is well known from discrete states.  
Im$(\omega)$ however  leads to
width bifurcation and finally (when the system is coupled to one channel,
i.e. to  a common continuum of scattering wavefunctions),  to a 
splitting of the system into two
parts with different characteristic time scales.
Physically, the system is stabilized by this splitting since the
lifetimes of most ($N-1$) states are longer than before while
that of only one state is shorter. In the cross section the
short-lived  state appears as a background term in high-resolution 
experiments. 
The wavefunctions of the long-lived states are mixed in those of the 
original ones in a comparably large parameter range.
Numerical results for the eigenvalues and eigenfunctions 
are shown for $N=2, ~4$ and 10 states coupled mostly to 1 channel.
\end{abstract}


\vspace{1cm}

\maketitle

\section{Introduction}
\label{intr}

The description of quantum mechanical systems by means of the
Schr\"odinger equation has been developed more than 80 years ago. At that
time only a few resonance states in nuclei and atoms were known which
are well separated from one another. The energies of these states are well 
described by means of
the Schr\"odinger equation with a Hermitian Hamiltonian.
In order to describe the finite lifetimes of these states, the R-matrix
theory has been developed which is however too complicated for
practical calculations. The finite lifetimes of the individual states
are calculated usually perturbatively, see e.g. \cite{rmatr}. 
Later, systems at high level density were in the center of
interest. Under these conditions, a statistical description of the
system turned out to be very efficient \cite{weidenmueller}. 
The resonances are well isolated from one another, and the lifetimes 
of the states are very long and do not play any role. 

In the course of time, experimental studies have been performed 
for different quantum systems with a much improved accuracy. 
Also the theoretical calculations are carried out
today not only for states being well separated in energy 
from one another but also for resonance 
states in the regime of overlapping. Here the 
single individual states can no longer be identified what results in 
problems of their interpretation. 
Contradictions between experimental results and conventional Hermitian
quantum physics appeared in different small quantum systems. For
example, Heiblum et al. \cite{laps1} found  experimentally a
crossover from the mesoscopic to a universal
phase  for electron transmission in quantum dots about than 10 years
ago. These results could not be explained in conventional quantum physics in
spite of much effort \cite{laps2}. 
Recently Koehler et al. \cite{koehler} observed that neutron resonance data
exclude random matrix theory. Deviations between experimental  data
and random matrix theory in nuclear physics studies were observed also 
earlier, e.g. \cite{kanter}. These and other experimental results
show that the Schr\"odinger equation  originally introduced with a
Hermitian Hamilton operator for the description of well isolated
resonances, has to be  extended. Above all,
the lifetimes of the resonance states have to be calculated also in  
the regime of overlapping resonances and the justification of
statistical assumptions has to be proven for small quantum systems.

The lifetimes of the resonance states can be calculated when 
the system is explicitly considered to be open and the calculations
are performed quantum mechanically for both the system and the
environment of scattering wavefunctions into which the system is embedded. 
Using the Feshbach projection operator
technique \cite{feshbach}, first the {\it energy-independent} many-body problem
of the system (with the Hermitian Hamiltonian
$H_B$) is solved in the standard manner.  
These solutions provide the energies $E_i^B$
and wavefunctions $\Phi_i^B$ of the discrete 
states with inclusion of the so-called  {\it internal} interaction. 
In a second step, the {\it energy-dependent} scattering wavefunctions
$\xi_c^E$ of the  
environment are calculated and, further, the (energy-dependent) 
coupling matrix elements  
\begin{eqnarray}
\gamma_{k c}^0  = 
\sqrt{2\pi}\, \langle \Phi_k^B| V | \xi^{E}_{c}
\rangle \; .
\label{form3}
\end{eqnarray}
between the discrete states of the system and the environment 
are evaluated (see \cite{top}, section 2.1). 
The corresponding Schr\"odinger equation 
contains an energy dependent (nonlinear)
source term describing the coupling between system and environment. 
The Hamiltonian ${\cal H}_0$ of this equation is non-Hermitian and 
provides the lifetimes of the states, but  without any additional
interaction of the states via the environment.
The Schr\"odinger equation  with  ${\cal H}_0$ and
source term can be rewritten into a
Schr\"odinger equation without source term but with a
non-Hermitian Hamiltonian ${\cal H}$ 
that contains the interaction of the states via
the environment (the so-called {\it external} interaction) 
in the nondiagonal matrix elements \cite{ro01}. The Hamiltonian $\ch$ reads
\begin{eqnarray}
\ch\; =\; H_{B} + V_{BC} G_C^{(+)} V_{CB} 
\label{form5}
\end{eqnarray}
where $V_{BC}$ and $V_{CB}$ stand for the coupling between system and
environment and
$G_C^{(+)}$ is the Green function in the subspace of scattering states.
The external interaction of the states via the continuum 
is complex, generally. The principal value integral is 
\begin{eqnarray}
{\rm Re}\; 
\langle \Phi_i^{B} | \ch |  \Phi_j^{B} \rangle 
 -  E_i^B \delta_{ij} 
=\frac{1}{2\pi} \sum_{c=1}^C {\cal P} 
\int_{\epsilon_c}^{\epsilon_{c}'} 
 {\rm d} E' \;  
\frac{\gamma_{ic}^0 \gamma_{jc}^0}{E-E'} 
\label{form11}
\end{eqnarray}
and the residuum reads
\begin{eqnarray}
{\rm Im}\; \langle \Phi_i^{B} | \ch |
  \Phi_j^{B} \rangle =
- \frac{1}{2}\; \sum_{c=1}^C  \gamma_{ic}^0 \gamma_{jc}^0 \; .
\label{form12}
\end{eqnarray}
The interaction of the states of the system via the environment 
is involved in the eigenvalues ${\cal E}_i$ and eigenfunctions 
$\Phi_i$ of the Hamiltonian  ${\cal H}$.
It is therefore relatively easy, in this formalism, 
to study the influence of the environment onto the states of the system.
That means, the non-Hermitian quantum physics is -- in contrast to
the widely spread meaning -- {\it not} a further
approximation introduced in the  theory. It is an expression of the
fact that the system considered is really open and its properties 
are influenced by the coupling to the environment.
This influence is  unimportant at low level density where it can be
described by perturbation theory. It becomes, however, decisive
in the regime of overlapping resonances.

Meanwhile, many calculations for open quantum systems are performed
with high accuracy. Without using any perturbation theory
or statistical assumptions, 
the Schr\"odinger  equation with the non-Hermitian Hamilton operator 
${\cal H}$ is solved and the eigenvalues ${\cal E}_i= E_i
-\frac{i}{2}\Gamma_i$ and eigenfunctions $\Phi_i$ are obtained. That means,
not only the energies $E_i$ of the states of the system 
are evaluated but also their lifetimes $\tau_i \propto 1/\Gamma_i$.
The control of the eigenvalues  by a parameter 
allows to draw conclusions on the dynamics of open quantum systems.

The limits of the applicability of the standard quantum
theory with Hermitian Hamilton operator
can be seen best at  high level density, where the
individual resonance states overlap and interact 
with one another via the continuum of scattering wavefunctions. 
Here, the dynamics of the system is determined by singular
points, the so-called {\it exceptional points}. They cause level
repulsion in energy as well as a bifurcation of the widths (inverse
proportional to the lifetimes) of the states.
Results are obtained theoretically as
well as experimentally, which are counterintuitive at first glance\,:
due to width bifurcation,
coherent short-lived states are formed together with states 
that are almost decoupled from the continuum of scattering states
\cite{top,klro,fdp1}. An example is the above-mentioned
crossover from the mesoscopic to a universal phase in the transmission 
in quantum dots \cite{laps1} and its qualitative explanation on the
basis of a Schr\"odinger equation with non-Hermitian Hamilton operator
\cite{laps3}. This phenomenon is observed in different systems and is
called usually {\it dynamical phase transition} \cite{pastawski}.
The short-lived states are analog  to the coherent superradiant states 
considered by Dicke \cite{dicke} in 1954  as shown by means
of a toy model \cite{soze}.  Moreover, it could be shown  in an experiment 
on non-locally coupled pairs of quantum point contacts that
discrete states undergo a robust interaction that is achieved
by coupling them to each other through the continuum \cite{bird}. 
Most of these results are of high value
for fundamental questions of quantum mechanics as well as for applications.

It is the aim of the present paper to show some generic (mostly numerical)
results for open quantum systems 
in order to receive a deeper understanding
for the dynamical phase transitions occurring in the regime of
overlapping resonances. A toy model is used in order
to receive conclusions on the role  played by  exceptional points and
avoided level crossings at high level density
in open quantum systems. Of special interest are the 
effects arising from the  imaginary part (\ref{form12})
of the coupling term via the environment.
 
The calculations are performed with respectively two, four 
and ten resonance states coupled  mostly to one open decay channel 
(corresponding to the common  continuum of scattering wavefunctions)
under the condition that they  avoid crossing
or even cross in a certain  parameter range. 
The coupling of $N$ resonance states to $K<N$ channels corresponds to
the general situation of open quantum systems.
The formalism used in the calculations is the same  as that discussed in
our earlier paper \cite{fdp2}.
The results show very clearly that the imaginary part of the coupling
term between system and environment causes width bifurcation and,
finally, the formation of different time scales in the regime of
overlapping resonances. Due to  width bifurcation, the system 
splits into two parts that exist at different times. This process is
irreversible. The wavefunctions of the short-lived {\it aligned} states 
are mixed coherently in relation to the open decay channel
while those of the long-lived {\it trapped} states are mixed
incoherently such that the states are almost decoupled from the open 
decay channel.
  
The results confirm the characteristic features of the
dynamical phase transitions appearing in open
quantum systems at high level density. They show
that the number of states of the system is reduced during the
dynamical phase transition due to ejecting the aligned short-lived state
and, furthermore, that the wavefunctions  of the states before and
beyond the dynamical phase transition are completely different from 
one another. While the states of the
original system (without interaction via the continuum)
have individual spectroscopic features, those of the system consisting
of the long-lived states beyond the dynamical phase transition show
chaotic features. These results prove once more 
that the experimental results \cite{laps1} on  the crossover from the
mesoscopic to a universal phase for electron transmission
in quantum dots can be explained 
by means of the non-Hermiticity of the Hamiltonian. 
Additionally it should be mentioned here that the aligned state 
corresponds to the
Dicke superradiant state and the trapped states to the subradiant
states which are considered in optics, e.g. \cite{kaiser}. 
 
In section \ref{form}, we sketch the formalism used in the present
calculations.  For $N=2$ states with equal decay widths, the analytical
and numerical solutions of the problem show clearly the effect of width
bifurcation. Numerical results for eigenvalues and eigenfunctions 
obtained for $N=2$, 4 and 10 states are provided in sections \ref{2} to
\ref{v10}. The results are discussed in section \ref{disc} while
conclusions on the relation between exceptional points and width
bifurcation are drawn in the last section \ref{concl}.

\section{Formalism}
\label{form}

We consider an $N\times N$ matrix 
\begin{eqnarray}
{\cal H} = 
\left( \begin{array}{cccc}
\epsilon_{1}+\omega_{11}  & \omega_{12} & \ldots &\omega_{1N}   \\
\omega_{21} & \epsilon_{2}+\omega_{22}  &  \ldots & \omega_{2N}\\
\vdots     & \vdots &             \ddots&   \vdots \\
\omega_{N1} & \omega_{N2}       &    \ldots   &  \epsilon_{N}+\omega_{NN} \\
\end{array} \right) 
\label{form1}
\end{eqnarray}
the diagonal elements of which are the $N$ complex eigenvalues 
$ \epsilon_{i} + \omega_{ii}
\equiv e_i - i/2~\gamma_i$ of a non-Hermitian operator. 
The $\omega_{ii}$ are the so-called selfenergies of the states
arising from their coupling  to the environment of scattering
wavefunctions into which the system is embedded.
In atomic physics, these values are  known as Lamb shift.
Our calculations are performed with  coupling matrix elements
$\omega_{ii}$ the values of which do not depend on the parameter considered. 
In such a case, the $\omega_{ii}$ can 
considered   to be included into the diagonal matrix elements,
which read  $\varepsilon_i \equiv \epsilon_{i}+\omega_{ii}$.
The $e_i$ and  $\gamma_i$ denote the 
energies and widths, respectively, of the $N$ states (including their
selfenergies) without account of the interaction 
of the different states via the environment.

The internal interaction of the two states $i$ and $k\ne i$
(appearing in the closed system) {\it as well as} their external interaction
(via the environment) are contained in the $\omega_{ik}$. 
The internal interaction can be caused only by some part of
Re$(\omega_{ik})$ while the external interaction contains complex 
$\omega_{ik}$, see equations (\ref{form11}) and (\ref{form12}). 
Most interesting part of the external interaction is therefore 
Im$(\omega_{ik})$. It becomes  important at high level
density where the corresponding resonance states overlap.

When the number $N$ of states is equal to the number $K$ of
common channels
(i.e. equal to the number $K$ of different common continua of scattering
wavefunctions) all the coupling matrix elements $\omega_{ik}$ are
different from zero and the matrix (\ref{form1}) is full.
In the case with only one open decay channel $K=1$, all
$\omega_{ik}$ different from
$\omega_{i~k=K}$ and  $\omega_{i=K~k}$ are zero. An example of $N=4$
states coupled to only the fourth channel is the following matrix
\begin{eqnarray}
{\cal H} = 
\left( \begin{array}{cccc}
\varepsilon_{1}  & 0 & 0 &\omega_{14}   \\
0 & \varepsilon_{2}   &  0 & \omega_{24}\\
0    & 0 &       \varepsilon_{3}    &   \omega_{34} \\
\omega_{41} & \omega_{42}       &   \omega_{43}  &  \varepsilon_{4} \\
\end{array} \right)
\label{int2}
\end{eqnarray}
with $\varepsilon_i = \epsilon_i$ for $i\ne 4$ and 
$\varepsilon_4 = \epsilon_4+\omega_{44}$.

The eigenvalues of $\ch$ will be denoted by $\ce_i 
\equiv E_i - i/2 ~\Gamma_i$ where  $E_i$ and $\Gamma_i$ stand for the
energy and width, respectively, of the eigenstate $i$. 
The eigenfunctions of the non-Hermitian
$\ch$ are biorthogonal (see sections 2.2 and 2.3 of \cite{top}), 
\begin{eqnarray}
\langle \Phi_i^*|\Phi_j\rangle = \delta_{ij} \; .
\label{int3}
\end{eqnarray}
It follows 
\begin{eqnarray}
 \langle\Phi_i|\Phi_i\rangle & = & 
{\rm Re}~(\langle\Phi_i|\Phi_i\rangle) ~; \quad
A_i \equiv \langle\Phi_i|\Phi_i\rangle \ge 1
\label{int4} 
\end{eqnarray}
and 
\begin{eqnarray}
\langle\Phi_i|\Phi_{j\ne i}\rangle & = &
i ~{\rm Im}~(\langle\Phi_i|\Phi_{j \ne i}\rangle) =
-\langle\Phi_{j \ne i}|\Phi_i\rangle 
\nonumber  \\
&& |B_i^j|  \equiv 
|\langle \Phi_i | \Phi_{j \ne i}| ~\ge ~0  \; .
\label{int5}
\end{eqnarray}
At an exceptional point, $A_i \to \infty$ and $|B_i^j| \to \infty$.
The $\ce_i$ and $\Phi_i$ contain global features that are 
caused by many-body forces  induced by the coupling
$\omega_{ik}$ of the states $i$ and $k\ne i$ via the environment, see 
equations (\ref{form11}), (\ref{form12}) and the corresponding discussion in
\cite{fdp1,jung}.

In the case $N=2$, the two eigenvalues of ${\cal H}$ are
\begin{eqnarray}
\ce_{i,j} \equiv E_{i,j} - \frac{i}{2} \Gamma_{i,j} = 
 \frac{\varepsilon_1 + \varepsilon_2}{2} \pm Z; \quad
Z \equiv \frac{1}{2} \sqrt{(\varepsilon_1 - \varepsilon_2)^2 + 4 \omega^2}
\; .
\label{int6}
\end{eqnarray}
According to this expression, two interacting discrete states (with
$\gamma_k = 0$) avoid always crossing since $\omega$ and 
$\varepsilon_1 - \varepsilon_2$ are real in this case. 
Resonance states with nonvanishing widths $\Gamma_i$ 
repel each other in energy  according to the value of Re$(Z) $
while the widths bifurcate according to the value of Im$(Z)  $.
The two states cross when $Z=0$. This crossing point is an exceptional
point according to the definition of Kato \cite{kato}.

At the exceptional point $Z=0$, the eigenfunctions of (\ref{form1})
of the two crossing states are linearly dependent from one another, 
\begin{eqnarray}
\Phi_1^{\rm cr} \to ~\pm ~i~\Phi_2^{\rm cr} \; ;
\quad \qquad \Phi_2^{\rm cr} \to
~\mp ~i~\Phi_1^{\rm cr}   
\label{eif5}
\end{eqnarray}  
according to analytical  as well as numerical and experimental
studies, see  Appendix of \cite{fdp1} and section 2.5 of \cite{top}.
That means, the wavefunction $\Phi_1$ of the state $1$ jumps, 
at the exceptional point, via the wavefunction ~$\Phi_1\pm \, i\, \Phi_2$  
of a chiral state to   ~$\pm\, i\, \Phi_2$
\cite{comment}. From (\ref{eif5}) follows\,:
\begin{verse}
(i) When  two levels are distant from one another,  their eigenfunctions
 are (almost) orthogonal,  
$\langle \Phi_k^* | \Phi_k \rangle   \approx
\langle \Phi_k | \Phi_k \rangle  = A_k \approx 1 $.\\
(ii) When  two levels cross at the exceptional point, 
their eigenfunctions are linearly
dependent according to (\ref{eif5}) and 
$\langle \Phi_k | \Phi_k \rangle \equiv A_k \to \infty $.\\
\end{verse}
These two relations show that the phases of the two eigenfunctions
relative to one another change when the crossing point is approached. 
This can be expressed quantitatively by defining the {\it phase
  rigidity} $r_k$ of the eigenfunctions $\Phi_k$,
\begin{eqnarray}
r_k ~\equiv ~\frac{\langle \Phi_k^* | \Phi_k \rangle}{\langle \Phi_k 
| \Phi_k \rangle} ~= ~A_k^{-1} \; . 
\label{eif11}
\end{eqnarray}
It holds $1 ~\ge ~r_k ~\ge ~0 $.  
The  non-rigidity $r_k$ of the phases of the eigenfunctions of $\ch$ 
follows also from the fact that $\langle\Phi_k^*|\Phi_k\rangle$
is a complex number (in difference to the norm
$\langle\Phi_k|\Phi_k\rangle$ which is a real number) 
such that the normalization condition
(\ref{int3}) can be fulfilled only by the additional postulation 
Im$\langle\Phi_k^*|\Phi_k\rangle =0$ (what corresponds to a rotation). 

When $r_k<1$, an analytical expression for the eigenfunctions as a
function of a certain control parameter  can,
generally, not be obtained. An exception is the special case  
$\gamma_1 = \gamma_2$  for which
$Z=  \frac{1}{2} \sqrt{(e_1 - e_2)^2 + 4 \omega^2}$. 
In this case, the condition $Z=0$ can not be fulfilled if 
$\omega = x$ is real due to 
\begin{eqnarray}
(e_1 - e_2)^2 +4\, x^2 &>& 0  \, .
\label{int6a}
\end{eqnarray}
The exceptional point can be found only by
analytical continuation into the continuum \cite{top,ro01} and  
the two states avoid crossing. This is analogue
to the avoided level crossings of discrete states. 

The condition $Z=0$ can be fulfilled however if
 $\omega = i\, x$ is imaginary,
\begin{eqnarray}
(e_1 - e_2)^2 -4\, x^2 &= &0 
~~\rightarrow ~~e_1 - e_2 =\pm \, 2\, x \; ,
\label{int6b}
\end{eqnarray}
and two exceptional points appear.  It holds further
\begin{eqnarray}
\label{int6c}
(e_1 - e_2)^2 >4\, x^2 &\rightarrow& ~Z ~\in ~\Re \\
\label{int6d}
(e_1 - e_2)^2 <4\, x^2 &\rightarrow&  ~Z ~\in ~\Im 
\end{eqnarray}
independent of the parameter dependence $e_i(a)$.
In the first case, the eigenvalues ${\cal E}_i = E_i-i/2\, \Gamma_i$ 
differ from the original values 
$\varepsilon_i = e_i - i/2~\gamma_i$ by a contribution to
the energies and in the second case by a contribution
to the widths. 
The width bifurcation starts at one of the exceptional points and
becomes maximum in the middle between the two exceptional points.
This happens at the crossing point $e_1 = e_2$ where 
$\Delta \Gamma/2 \equiv |\Gamma_1/2 - \Gamma_2/2| = 4\, x$.  

Some years ago, the case $N=2$ with $e_i=e_i(a)$, 
fixed real $\omega \equiv \omega_{12} = \omega_{21}$,
and different fixed values of $\gamma_i$, including $\gamma_i = 0$, 
is  studied  as a function of the parameter $a$ in the neighborhood 
of avoided and true
crossings of the two levels \cite{ro01}.  The results for the 
$N=2$ case  \cite{ro01} show further that the wavefunctions of the
two states $\Phi_1$ and $\Phi_2$ are mixed in a  finite range of the
parameter $a$ around the critical value $a_{\rm cr}$
at which the two states
avoid crossing. This holds true not only for resonance states but also
for discrete states. 

The  eigenfunctions $\Phi_i$ of ${\cal H}$ can be represented in the
set of basic wavefunctions $\phi_i$ of the unperturbed matrix
(corresponding to the case with vanishing coupling matrix elements
$\omega_{ij}$),
\begin{eqnarray}
\Phi_i=\sum_{j=1}^N b_{ij} \phi_j \; .
\label{int20}
\end{eqnarray}
Also the $b_{ij}$ are normalized  according to the biorthogonality
relations  of the wavefunctions $\{\Phi_i\}$. 

In our calculations, the mixing coefficients $b_{ij}$ 
of the wavefunctions of the two  states due to their avoided  
crossing are not calculated. We simulate the  fact  that  the two 
wavefunctions  are mixed in a finite parameter range around the
critical value of their avoid  crossing \cite{ro01} by assuming 
a  Gaussian distribution 
\begin{eqnarray}
\omega_{i\ne j} = \omega ~e^{-(e_i -e_j)^2}
\label{int7}
\end{eqnarray}
for the coupling coefficients. The results reproduce very well those
discussed in \cite{ro01} for 2 levels and real coupling $\omega$. 

Of special interest is the situation 
at high level density where the ranges of avoided crossings,
defined by (\ref{int7}),  of
different levels overlap.  Some generic results obtained  
with 2, 4 and 10 resonance states will be presented in the following 
sections.

\section{Eigenvalues and eigenfunctions for ${\bf \boldmath N=2}$ 
crossing levels}
\label{2}

We start our calculations with the two-level case which is studied
mostly in literature. 
We choose the matrix (\ref{form1}) with  $e_i=e_i(a)$, 
different fixed values of $\gamma_i$ and  
fixed $\omega$.
The functional dependence of the energies over the parameter $a$ is 
similar as in \cite{ro01,fdp2}. However, the $\omega$ are 
real in \cite{ro01} while they are mostly complex
in the present paper as well as in \cite{fdp2}. According to
(\ref{int6}) to (\ref{int6d}), the exceptional points appear at 
different values of the 
parameter $a$  when the ratio Re$(\omega)$ to Im$(\omega)$ is varied, 
see also Figs. 1 to 3 in \cite{fdp2}.  

\begin{figure}[ht]
\begin{center}
%\includegraphics[width=6cm,height=3cm]{mixing-2_level_05a-gen.eps} a
%~~\includegraphics[width=6cm,height=3cm]{mixing-2_level_05b-gen.eps} b
%\\[.3cm] 
%\includegraphics[width=6cm,height=3cm]{mixing-2_level_06a-gen.eps} c
%~~\includegraphics[width=6cm,height=3cm]{mixing-2_level_06b-gen.eps} d
%\\[.3cm] 
%\includegraphics[width=6cm,height=3cm]{mixing-2_level_04a-gen.eps} e
%~~\includegraphics[width=6cm,height=3cm]{mixing-2_level_04b-gen.eps} f
%\\[.3cm] 
%\includegraphics[width=6cm,height=3cm]{mixing-2_level_07a-gen.eps} g
%~~\includegraphics[width=6cm,height=3cm]{mixing-2_level_07b-gen.eps} h
\includegraphics[width=14cm,height=14cm]{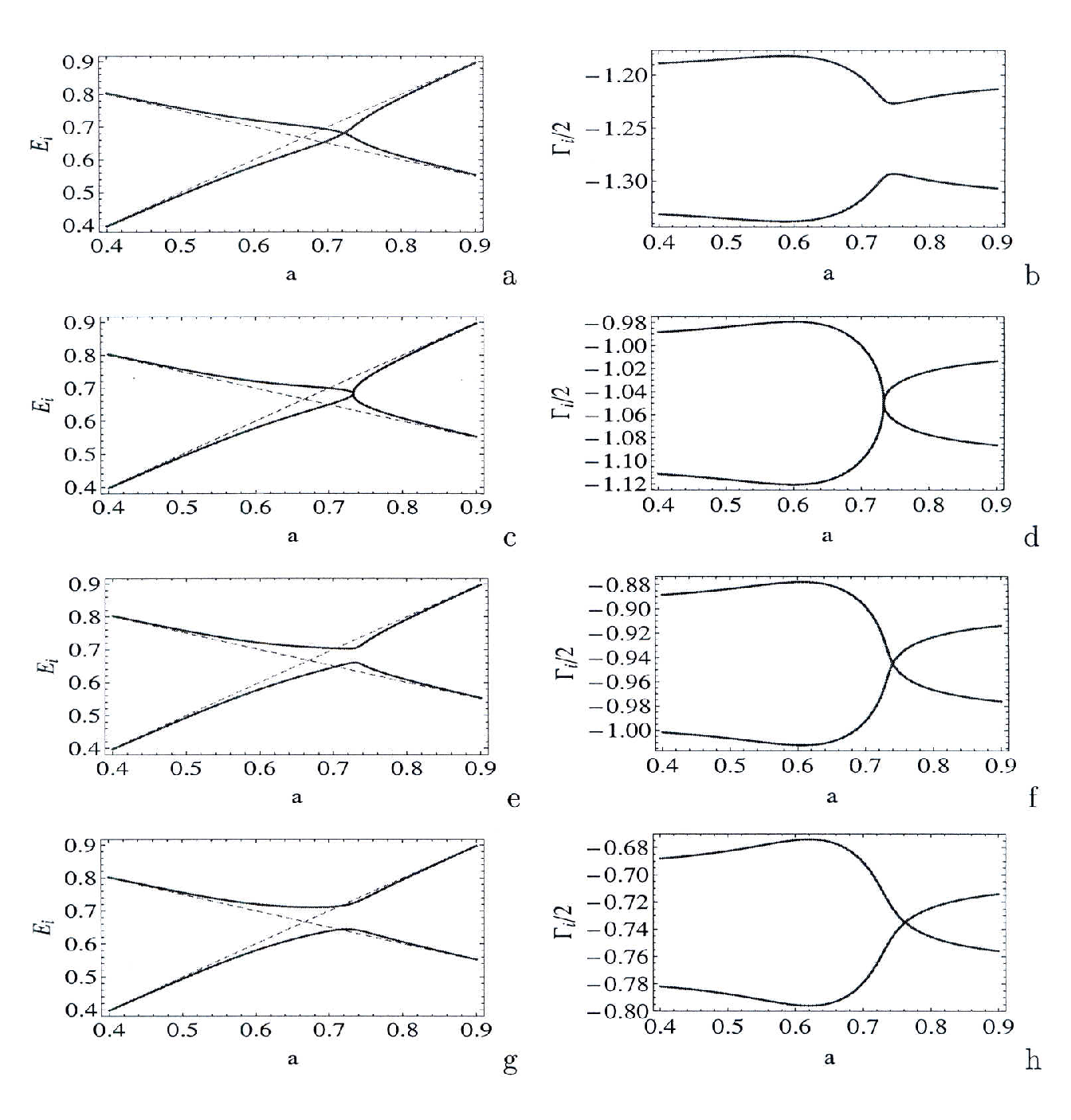}
\end{center}
\caption{\footnotesize
Energies $E_i$ and widths $\Gamma_i/2$ (full lines) 
of $N=2$ states beyond ~(a,b), at
~(c,d) and below ~(e to h) the exceptional point. 
The parameters of the subfigures 
are $\gamma_1/2 = 1.2$ ~(a,b); ~$\gamma_1/2 =  1.0$ ~(c,d);
~$\gamma_1/2 =  0.9 $ ~(e.f); ~$\gamma_1/2 =  0.7 $ ~(g,h). 
Further parameters:
~$e_1=1-a/2; ~e_2=a; ~\gamma_2/2 = 1.1 \gamma_1/2;
~\omega = (1+i)~0.05$. The dashed lines show $e_i(a)$. 
}
\label{fig1}
\end{figure}

In Fig. \ref{fig1} we show the numerical results for the avoided level
crossing phenomenon as a function of the parameter $a$
beyond, at and below the exceptional point. In all cases, 
the value of $\omega$ is fixed to $0.05 ~(1+i)$ while the  
$\gamma_1/2$ are different in the different subfigures (with
$\gamma_2/\gamma_1= 1.1$ in all cases).
The exceptional point appears at $\gamma_1^{\rm cr}/2 = 1$, see
Figs. \ref{fig1}.\,c, d. When $\gamma_1/2 >  \gamma_1^{\rm cr}/2$
(Figs. \ref{fig1}.\,a, b), the eigenvalue trajectories cross  in
energy while  the widths $\Gamma_1$ and  $\Gamma_2$ are always
different from one another. The situation is another one when 
$\gamma_1/2 <  \gamma_1^{\rm cr}/2$. Here, the eigenvalue trajectories
avoid crossing in energy while the trajectories $\Gamma_1$ and  $\Gamma_2$
cross at certain values of $a$ (Figs. \ref{fig1}.\,e to h). 

The comparison of Fig. \ref{fig1} with complex $\omega =0.05~(1+i)$ and 
Fig. 2 in \cite{ro01} with real $\omega = 0.05$ shows the influence of  
Im$(\omega)$. While the $\Gamma_i(a)$ vary symmetrically around the
critical value $a^{\rm cr}$ when $\omega$ is real, this is not the case when 
$\omega$ is complex. In the last case, the difference $\Gamma_1 - \Gamma_2$
blows up in the critical region. Furthermore, the $\Gamma_i$
approach their asymptotic values in a relatively small parameter range
of $a$ when $\omega$ is real in contrast to the case with complex $\omega$.
These results will be discussed further in the following sections.

\begin{figure}[ht]
\begin{center}
%\includegraphics[width=6cm,height=3cm]{mixing-2_level_05c-new.eps} a
%\\[.3cm] 
%\includegraphics[width=6cm,height=3cm]{mixing-2_level_06c-new.eps} b
%\\[.3cm] 
%\includegraphics[width=6cm,height=3cm]{mixing-2_level_04c-new.eps} c
%\\[.3cm] 
%\includegraphics[width=6cm,height=3cm]{mixing-2_level_07c-new.eps} d
\includegraphics[width=7cm,height=14cm]{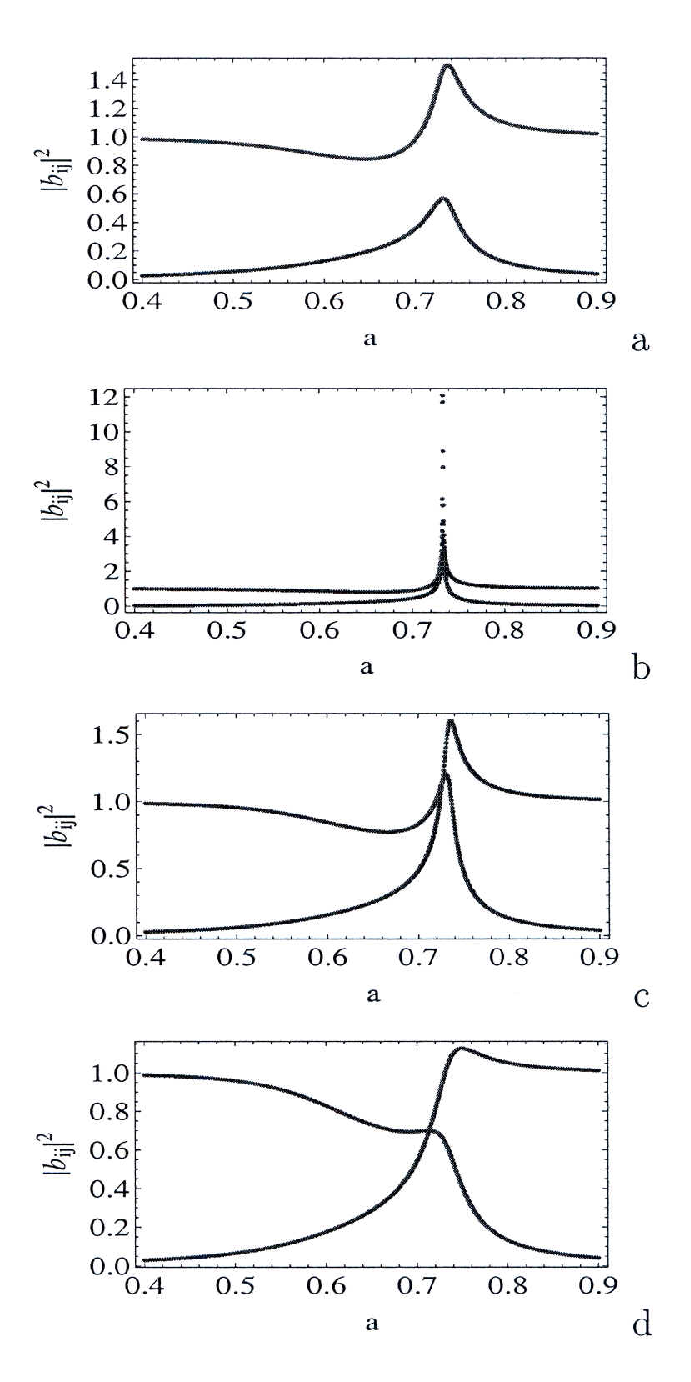}
\end{center}
\caption{\footnotesize
The mixing coefficients $|b_{ij}|^2$ of $N=2$ states beyond (a), at
(b) and below (c,d) the exceptional point. The parameters are the
same as in Fig. \ref{fig1}.
}
\label{fig2}
\end{figure}

In Fig. \ref{fig2}, we show the mixing coefficients $|b_{ij}|^2$ of the
eigenfunctions as a function of the parameter $a$
corresponding to the eigenvalue figures shown in
Fig. \ref{fig1}. The mixing coefficient is complex, $\omega = 0.05~(1+i)$.
This figure can be compared with Fig. 5 in \cite{ro01} calculated 
exactly with real $\omega = 0.05$.
In both cases, the difference between  the two curves $|b_{ij}|^2$
is always 1 when $\gamma_1/2 >  \gamma_1^{\rm cr}/2$. This result
corresponds to the fact that the 
two states will not be  exchanged when $\gamma_1/2 >  \gamma_1^{\rm cr}/2$.
The energy trajectories of the
eigenstates cross freely and the widths of the two states are always
different from one another, see Figs. \ref{fig1}.\,a, b.

At the exceptional point (where $\gamma_1/2 = \gamma_1^{\rm cr}/2$),
$|b_{ij}|^2 \to \infty$ in both cases.
When $\gamma_1/2 <  \gamma_1^{\rm cr}/2$, the two curves
$|b_{ij}|^2$ coincide at a
certain value of $a$. This happens for real (see \cite{ro01})
as well as for complex (Fig. \ref{fig2})
$\omega$, but is symmetrically only for real   $\omega$. At these
points, the two states are exchanged.  

In both cases $\gamma_1/2 <  \gamma_1^{\rm cr}/2$ and 
$\gamma_1/2 >  \gamma_1^{\rm cr}/2$, the asymptotic values 
1 and 0 are reached in a smaller parameter range when $\omega$ is real
than in the case with complex   $\omega$.
Within this parameter range the two eigenfunctions are mixed.  
The range of mixing  shrinks to one point when the two states really
cross (at the exceptional point). It is especially large when 
two discrete states (with $\gamma_i=0$) avoid crossing,
as can be seen from Fig. 5 in \cite{ro01} (where the calculations are 
performed exactly with real $\omega$ for all $b_{ij}$). 

The parameter range in which two states are mixed due to the
existence of an exceptional point  in the neighborhood 
plays an important role in realistic systems. It will be discussed
in the following sections with more than one avoided crossing
for both real and complex $\omega$.

\section{Eigenvalues for  ${\bf \boldmath N=4}$  crossing levels}
\label{v4}

\begin{figure}[ht]
\begin{center}
%\includegraphics[width=6cm,height=3cm]{fi3a-gen.eps} a
%~~\includegraphics[width=6cm,height=3cm]{fi3b-gen.eps} b
%\\[.4cm] 
%\includegraphics[width=6cm,height=3cm]{Mixing-053a-gen.eps} c
%~~\includegraphics[width=6cm,height=3cm]{Mixing-053b-gen.eps} d
%\\[.4cm] 
%\includegraphics[width=6cm,height=3cm]{Mixi-1a-gen.eps} e
%~~\includegraphics[width=6cm,height=3cm]{Mixi-1b-gen.eps} f
%\\[.4cm] 
%\includegraphics[width=6cm,height=3cm]{fi4a-gen.eps} g
%~~\includegraphics[width=6cm,height=3cm]{fi4b-gen.eps} h
%\\[.4cm] 
%\includegraphics[width=6cm,height=3cm]{Mixing-054a-gen.eps} i
%~~\includegraphics[width=6cm,height=3cm]{Mixing-054b-gen.eps} j
\includegraphics[width=13cm,height=17cm]{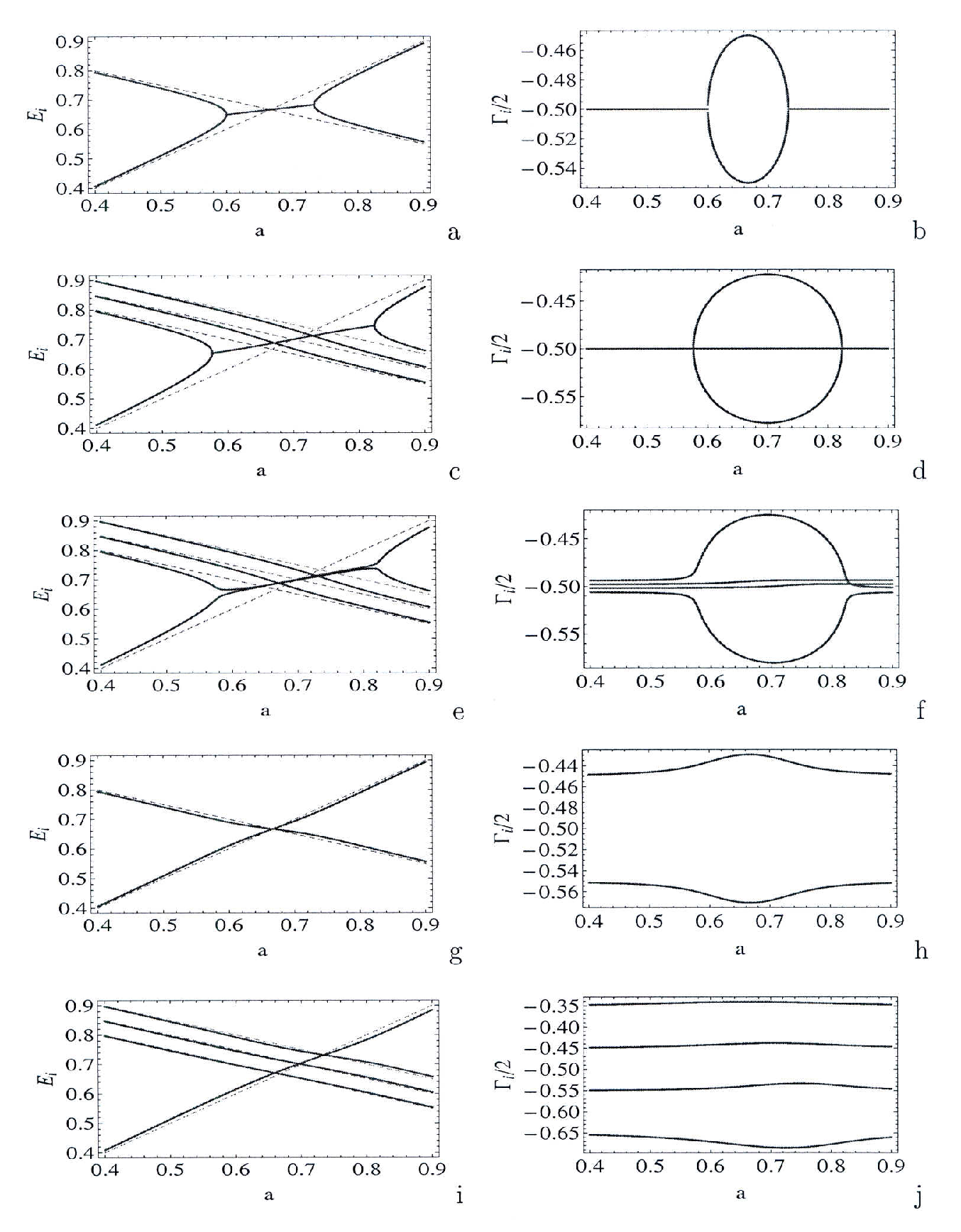}
\end{center}
\caption{\scriptsize
Energies $E_i$ and widths $\Gamma_i/2$ (full lines) 
of $N=2$ states (a,b,g,h) and $N=4$ states (c to f and i, j) coupled to
$K=1$ channel.
The parameters of the subfigures are 
$\gamma_i/2 = 0.5$ (a to d); ~$\gamma_i/2 = ~0.494, ~0.498, ~0.502,
~0.506$ (e,f); ~$\gamma_1/2 = 0.45, ~\gamma_2/2 = 0.55 $ (g,h); 
~$\gamma_1/2 = 0.35; ~\gamma_2/2 = 0.45; ~\gamma_3/2 = 0.55; 
~\gamma_4/2 =0.65$ (i,j).
Further parameters: $N=2:
~e_1=1-a/2; ~e_2=a ~; ~N=4: ~e_1=1-a/2; ~e_2=1.05-a/2; ~e_3=1.1-a/2;
~e_4=a; ~\omega = 0.05~i$. The dashed lines show $e_i(a)$. 
}
\label{fig3}
\end{figure}

\begin{figure}[ht]
\begin{center}
%\includegraphics[width=6cm,height=3cm]{Mixing-011a-gen.eps} a
%~~\includegraphics[width=6cm,height=3cm]{Mixing-011b-gen.eps} b
%\\[.4cm]
%\includegraphics[width=6cm,height=3cm]{Mixing-012a-gen.eps} c
%~~\includegraphics[width=6cm,height=3cm]{Mixing-012b-gen.eps} d
%\\[.4cm]
%\includegraphics[width=6cm,height=3cm]{Mixing-013a-gen.eps} e
%~~\includegraphics[width=6cm,height=3cm]{Mixing-013b-gen.eps} f
%\\[.4cm]
%\includegraphics[width=6cm,height=3cm]{Mixing-014a-gen.eps} g
%~~\includegraphics[width=6cm,height=3cm]{Mixing-014b-gen.eps} h
\includegraphics[width=14cm,height=14cm]{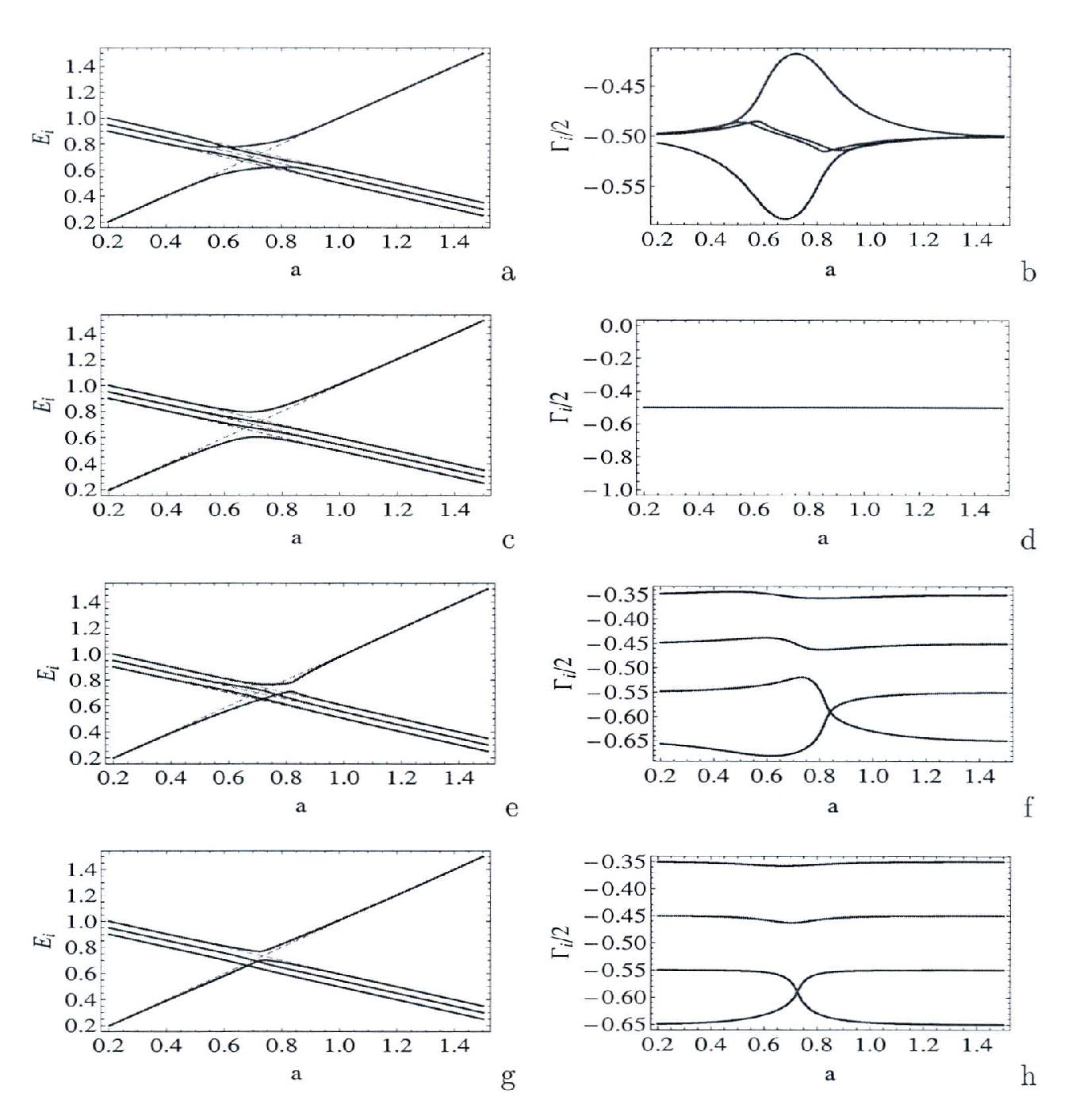}
\end{center}
\caption{\footnotesize
The same as Fig. \ref{fig3} but $\omega = (1+i)~0.05$ (a,b,e,f);
$\omega = 0.05$ (c,d,g,h); ~$N=4$; ~$\gamma_i= 0.5$ (a to d);
~$\gamma_i= 0.35; ~0.45; ~0.55; ~0.65$ (e to h).
}
\label{fig4}
\end{figure}

\begin{figure}[ht]
\vspace*{.7cm}
\begin{center}
%\includegraphics[width=6cm,height=3cm]{Mixing-022a-gen.eps} a
%~~\includegraphics[width=6cm,height=3cm]{Mixing-022b-gen.eps} b
%\\[.3cm] 
%\includegraphics[width=6cm,height=3cm]{Mixing-42a-gen.eps} c
%~~\includegraphics[width=6cm,height=3cm]{Mixing-42b-gen.eps} d
%\\[.3cm] 
%\includegraphics[width=6cm,height=3cm]{Mixing-32a-gen.eps} e
%~~\includegraphics[width=6cm,height=3cm]{Mixing-32b-gen.eps} f
\includegraphics[width=14cm,height=10.5cm]{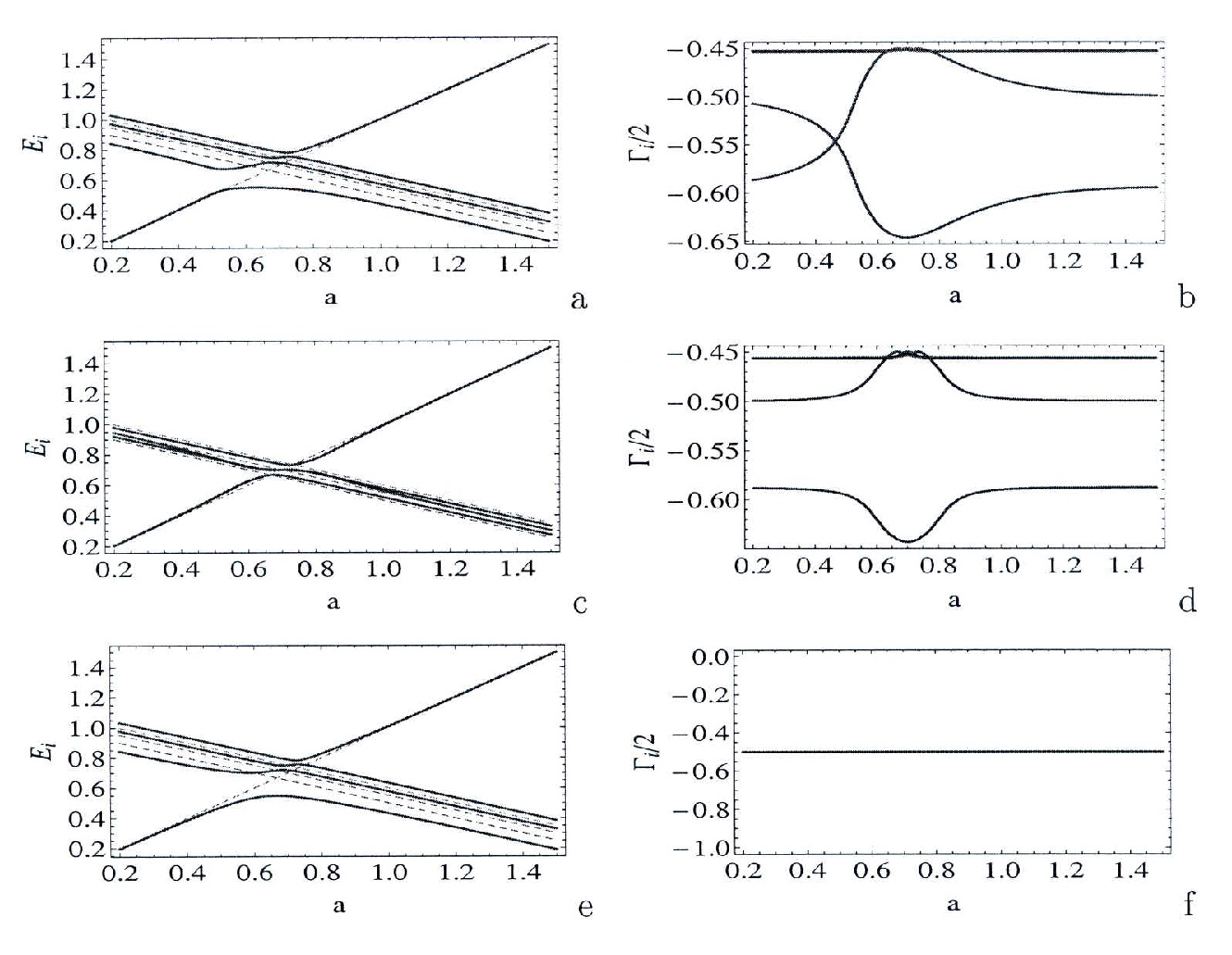}
\end{center}
\caption{\footnotesize
The same as Fig. \ref{fig3} but $N=4$ and $K=4$; 
~$\omega = (1+i)~0.05$ ~(a,b);
~$\omega = 0.05~i$ ~(c,d); ~$\omega = 0.05$ ~(e,f);
~$\gamma_i/2=0.5$.\\
}
\label{fig5}
\end{figure}

Let us first compare the calculations with $N=2$ and $N=4$ states 
and imaginary coupling $\omega = 0.05~i$ between discrete states and
environment of scattering states (Fig. \ref{fig3}). 
In the calculations, the eigenvalue trajectories $E_i(a)$ and $\Gamma_i(a)$
are traced as a function of the parameter $a$ which is varied in
such a manner that one state crosses in energy one and three,
respectively, other states. The results show two exceptional points 
in both cases appearing at the two
different parameter values $a^{\rm cr}_1$ and $a^{\rm cr}_2$, see
Figs. \ref{fig3}.a, b for two states and Figs. \ref{fig3}.c to f for 
four states. In between these two values, the
widths bifurcate\,: the differences $\Delta \Gamma/2 \equiv |\Gamma_1/2 -
\Gamma_2/2|$  blow up.  The two-level case is described 
analytically by equations (\ref{int6}) to (\ref{int6d}). 
The width bifurcation is completely reproduced in the numerical results.

The parameter range $|a^{\rm cr}_1 - a^{\rm cr}_2|$ is larger in the
case with four states than in the other one due to the larger region
in which avoided level crossings take place. Furthermore, the difference
$\Delta \Gamma/2$  is larger in the case of 4 states than in the case of 2
states. In the case that all states have equal widths $\gamma_i/2$,
the widths $\Gamma_i/2$ of only two states bifurcate also in the 
case with four states. The widths of the two other  states  remain unchanged   
(Figs. \ref{fig3}.c, d). This result holds also when the different 
$\gamma_i/2$ differ slightly from one another (Figs. \ref{fig3}.e, f). 
In this latter case, the two states avoid crossing at the two critical
values $a^{\rm cr}_1$ and $a^{\rm cr}_2$ and the width bifurcation is
almost the same as in the case with equal widths $\gamma_i$.

The situation changes completely when the widths $\gamma_i$ differ
stronger from one another. When $|\gamma_i/2 - \gamma_{i\pm 1}/2| >
|\epsilon_i - \epsilon_{i\pm 1}|$, the states do not avoid crossing
neither in the case with two  states nor in the case with four
states (Figs. \ref{fig3}.g to j). The eigenvalue trajectories cross
freely in energy  and the widths do (almost) not bifurcate.  
The dynamics of the system is therefore completely different from that
determined by the results shown in Figs. \ref{fig3}.a to f.

In Fig. \ref{fig4}, we show the eigenvalues $E_i$ and $\Gamma_i/2$ for
the case with four levels and one channel with complex and real coupling 
coefficients ($\omega = 0.05~(1+i)$ and   $\omega = 0.05$,
respectively). Width bifurcation  can be seen when 
$\omega$ is complex and the widths $\gamma_i$ are equal
(Figs. \ref{fig4}.a, b), 
while the $\Gamma_i = \gamma_i $ are independent of $a$  when $\omega$
is real (Figs. \ref{fig4}.c, d).
The results with different $\gamma_i$ differ from those obtained with
imaginary coupling $\omega = 0.05 ~i$ (Figs. \ref{fig3}.g to j)\,:
the widths of two states cross  at parameter values at which
their energies avoid crossing (Figs. \ref{fig4}.e to h).

In Fig. \ref{fig5}, we show the results of calculations with equal
number of states and channels, $N=4$ and $K=4$.  Although usually
$K<N$ in realistic systems, the results allow us to receive a deeper
understanding for the spectroscopic redistribution processes taking 
place in the critical region. The calculations are performed with the
same parametric dependence of the energies as in the foregoing
calculations with $N=4$ and $K=1$.  Further assumptions\,:
$\gamma_i=0.5$ and $\omega = 0.05~(1+i), ~~0.05 ~i$, and 0.05,
respectively.   

The results show width bifurcation when $\omega$ is complex or
imaginary (Figs. \ref{fig5}.a to d) which is however smaller than in
the corresponding cases with $K=1$ (Figs. \ref{fig4}.a, b and  
Figs. \ref{fig3}.c, d, respectively). Instead, the widths $\Gamma_i$
of the states are changed asymptotically due to the coupling to the
different channels.  When $\omega$ is real
(Figs. \ref{fig5}.e,f), the widths $\Gamma_i/2$ are
independent of $a$ and are equal to  $\gamma_i/2 = 0.05$ ~($i=$ 1 to 4).

\section{Eigenfunctions for  ${\bf \boldmath N=4}$  crossing levels}
\label{f4}

\begin{figure}[ht]
\begin{center}
\includegraphics[width=7cm,height=18cm]{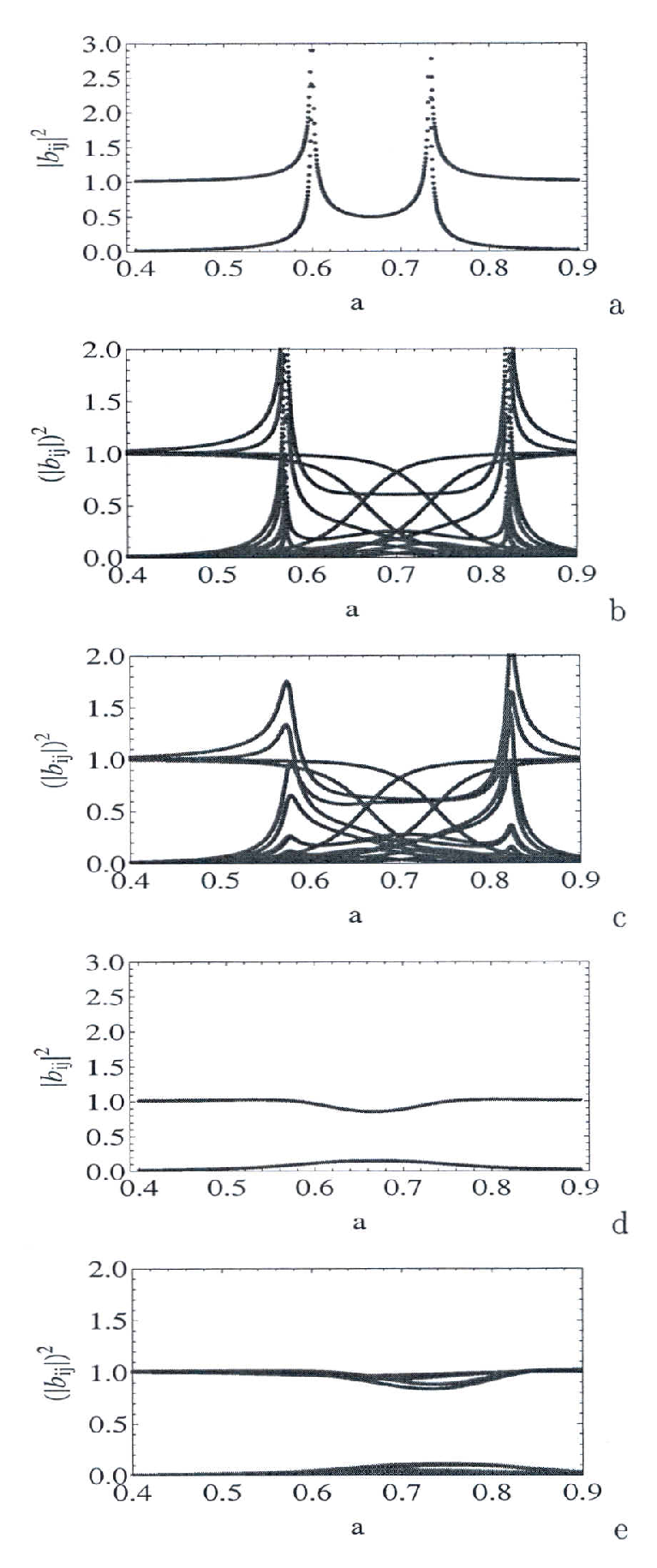}
\end{center}
\caption{\scriptsize
The mixing coefficients $|b_{ij}|^2$ of $N=2$ and $N=4$ states
the eigenvalues of which are shown in  Fig. \ref{fig3}; ~$K=1$ channel;
~$\omega = 0.05~i$.
}
\label{fig6}
\end{figure}

\begin{figure}[ht]
\begin{center}
\includegraphics[width=8cm,height=15.5cm]{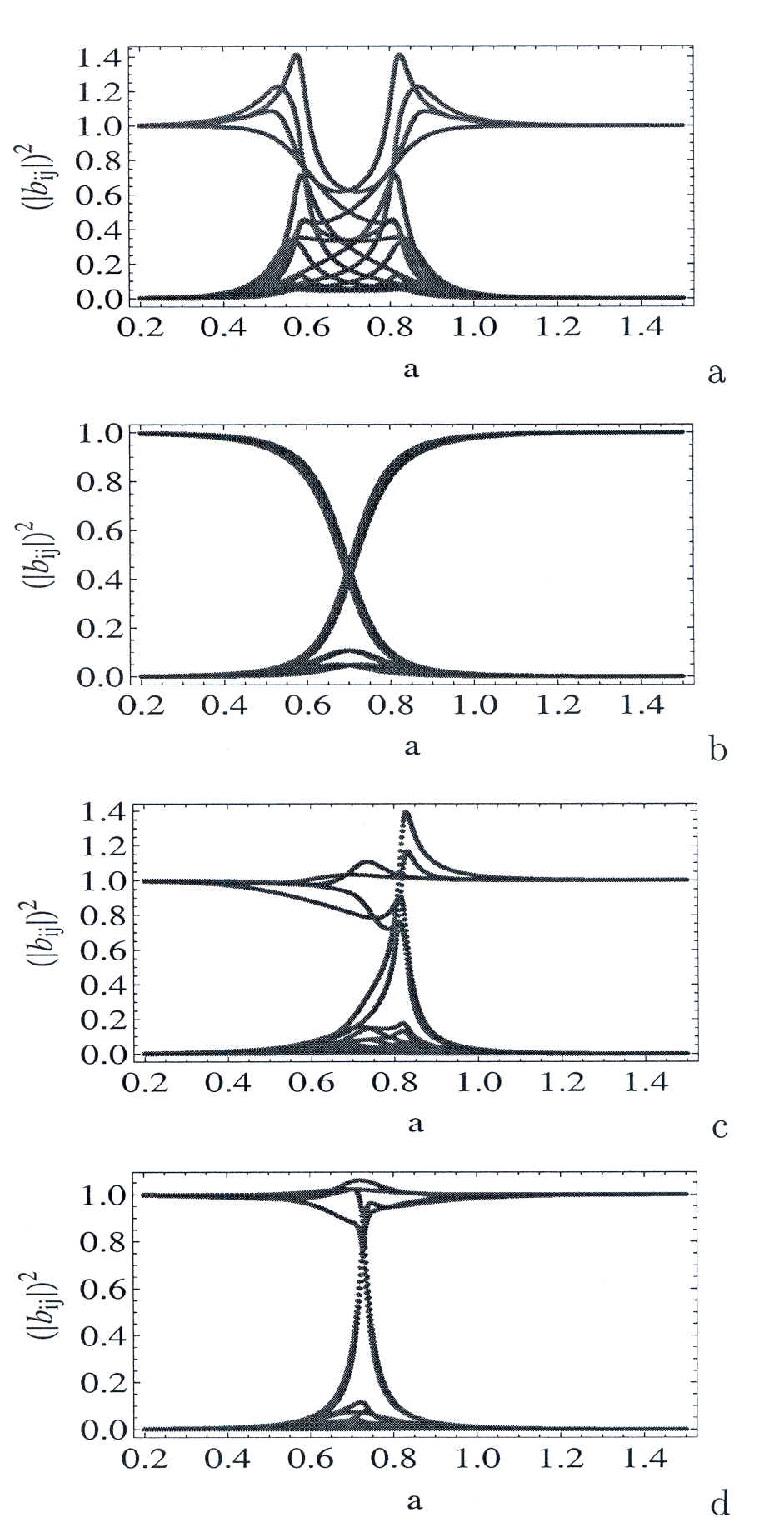}
\end{center}
\caption{\footnotesize
The mixing coefficients $|b_{ij}|^2$ of $N=4$ states
the eigenvalues of which are shown in  Fig. \ref{fig4}; $K=1$ channel;
~$\omega = (1+i)~0.05$ ~(a,c); ~$\omega = 0.05$ ~(b,d).
}
\label{fig7}
\end{figure}

\begin{figure}[ht]
\begin{center}
%\includegraphics[width=7cm,height=3.5cm]{Mixing-22c-new.eps} a
%\\[.4cm] 
%\includegraphics[width=7cm,height=3.5cm]{Mixing-42c-new.eps} b
%\\[.4cm] 
%\includegraphics[width=7cm,height=3.5cm]{Mixing-32c-new.eps} c
\includegraphics[width=8cm,height=12cm]{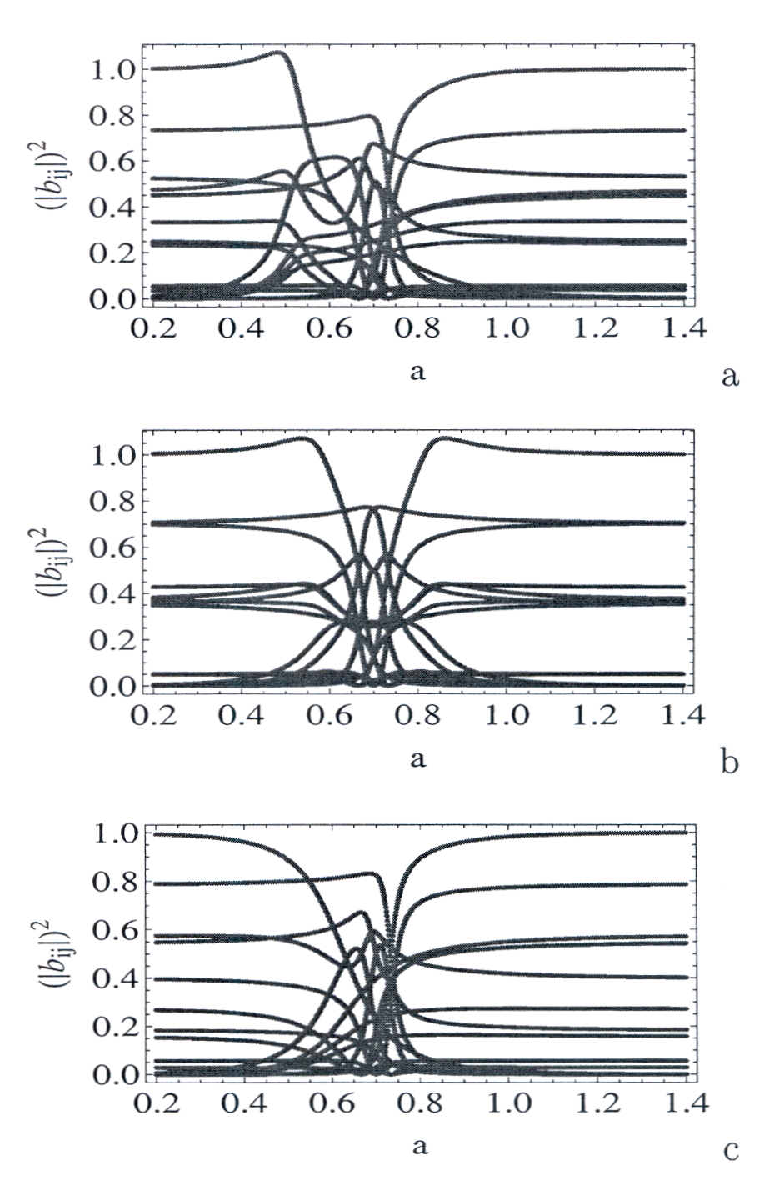}
\end{center}
\caption{\footnotesize
The mixing coefficients $|b_{ij}|^2$ of $N=4$ states
the eigenvalues of which are shown in  Fig. \ref{fig5}; $K=4$ channels;
~$\omega = (1+i)~0.05$ ~(a); ~$\omega = 0.05~i$ ~(b); ~$\omega = 0.05$ ~(c).
 \\
}
\label{fig8}
\end{figure}

The coefficients $b_{ij}$ defined in (\ref{int20}) determine the mixing
of the eigenfunctions $\Phi_i$ in relation to the basic wavefunctions
$\phi_j$. The mixing is nonvanishing  in the neighborhood of
the critical points of avoided level crossings as has been shown in
exact numerical calculations for two crossing states and real coupling 
coefficients $\omega_{ij}$ \cite{ro01}.
In the calculations of the present paper, the mixing of the
wavefunctions is simulated by assuming a Gaussian distribution for the
$\omega_{ij}$,  equation (\ref{int7}). It reproduces very well the exact results
obtained in \cite{ro01}. 

In the following, we consider 
the mixing coefficients $b_{ij}$ for all
wavefunctions $\Phi_i$ the eigenvalue trajectories of which are studied 
in Figs. \ref{fig3} to \ref{fig5} as a function of the parameter $a$.
In Fig. \ref{fig6}, the coefficients $|b_{ij}|^2$ are shown for two
and four states with imaginary coupling $\omega = 0.05 ~i$ (see the
corresponding eigenvalue trajectories in Fig. \ref{fig3}). 
When the widths $\gamma_i$ of all states are equal
(Figs. \ref{fig6}.a, b),
$|b_{ij}|^2\to \infty$ at the two exceptional points   
$a^{\rm cr}_1$ and $a^{\rm cr}_2$. In the
parameter range $a^{\rm cr}_1 < a < a^{\rm cr}_2$, the wavefunctions
are mixed while  $b_{ii} \to 1$ and $b_{ij}\to 0$ for $i\ne j$ 
beyond the two values  $a^{\rm cr}_{1,2}$. 
The states are completely (1:1) mixed   
in the range $a^{\rm cr}_1 < a < a^{\rm cr}_2$
in the two-level-case (Fig. \ref{fig6}.a). The picture is more
complicated in the four-level-case (Fig. \ref{fig6}.b).
Here, all states are involved
in the redistribution taking place in the critical parameter range 
$a^{\rm cr}_1 < a < a^{\rm cr}_2$. The exchange of the two states
the widths of which remain unchanged according to Fig. \ref{fig3}.d
can be seen from the energy eigenvalue trajectories, Fig. \ref{fig3}.c, 
as well from the mixing coefficients,  Fig. \ref{fig6}.b.

The figures are similar when the widths $\gamma_i$ of the states
differ slightly from one another. Instead of exceptional points, there
are avoided level crossings at   $a^{\rm cr}_1$ and $a^{\rm cr}_2$,
see Fig. \ref{fig6}.c for the four-level-case. The mixing coefficients
show a  dependence on $a$ which is similar to that obtained for equal
widths (Fig. \ref{fig6}.b). 

When the widths $\gamma_i$ of the states $i$  differ stronger from one
another than in Fig.  \ref{fig6}.c, the eigenfunctions remain almost
unmixed for all parameter values $a$ (Figs. \ref{fig6}.d, e). 
This result corresponds to the almost constant width trajectories $\Gamma_i$
as a function of $a$ (Figs. \ref{fig3}.h, j). An exchange of states
does not take place (Figs. \ref{fig3}.g, i).
These results show the great influence of the exceptional points onto
the dynamics of the system considered. 

The calculations in Figs. \ref{fig3} and \ref{fig6} are performed with
imaginary coupling $\omega = 0.05 ~i$. For comparison, we show in 
Fig. \ref{fig7} the mixing coefficients  $|b_{ij}|^2$ when $\omega$ 
is complex and real, respectively, for the four-level-case
(corresponding to the eigenvalue trajectories in Fig. \ref{fig4}). 
When all $\gamma_i$ are equal to one another
and $\omega$ is complex, the mixing coefficients 
$|b_{ij}|^2$ point to the participation of all states in the 
redistribution process in the critical parameter range
(Fig. \ref{fig7}.a).  This corresponds fully to the corresponding 
eigenvalue trajectories that show level exchange as well as width
bifurcation (Figs. \ref{fig4}.a,b). When the coupling is real,
however, the wavefunctions of only two states are mixed while the 
wavefunctions of the other two states remain almost unchanged in the
critical region (Fig. \ref{fig7}.b).

The mixing of the eigenfunctions is completely different from that 
discussed above  when the widths $\gamma_i$ differ 
more strongly from one another. A
mixing of the wavefunctions appears only when the energy trajectories
$E_i(a)$ avoid crossing and the width trajectories $\Gamma_i(a)$ cross, compare
Figs. \ref{fig4}.e to h with the corresponding  Figs. \ref{fig7}.c, d.    
A width bifurcation does not take place.

The mixing coefficients $|b_{ij}|^2$ for the case with $N=4$ states
coupled to $K=4$ channels are shown in Fig. \ref{fig8} (parameters
the same as in Fig. \ref{fig5}). The differences between the cases
with coupling to 4 channels to those with coupling  to only 1 channel 
can be seen by comparing Fig. \ref{fig8}.a with  Fig. \ref{fig7}.a, 
~Fig. \ref{fig8}.b with  \ref{fig6}.b and Fig. \ref{fig8}.c with 
Fig. \ref{fig7}.b. The mixing of the wavefunctions by coupling the
system to four channels is more complicated than that by coupling to
only one channel. This statement holds true also when $\omega$ is
imaginary. Furthermore, an additional mixing is caused by the
avoided level crossing at small $a$ in the four-channel-case when
$\omega$ is complex.

\section{Eigenvalues for  ${\bf \boldmath N=10}$  crossing levels}
\label{v10}

\begin{figure}[ht]
\begin{center}
%\includegraphics[width=6cm,height=3cm]{Multilevel-69a-gen.eps} a
%~~\includegraphics[width=6cm,height=3cm]{Multilevel-69b-gen.eps} b
%\\[.5cm]
%\includegraphics[width=6cm,height=3cm]{Multilevel-16a-gen.eps} c
%~~\includegraphics[width=6cm,height=3cm]{Multilevel-16b-gen.eps} d
%\\[.5cm]
%\includegraphics[width=6cm,height=3cm]{Multilevel-17a-gen.eps} e
%~~\includegraphics[width=6cm,height=3cm]{Multilevel-17b-gen.eps} f
\includegraphics[width=14cm,height=11cm]{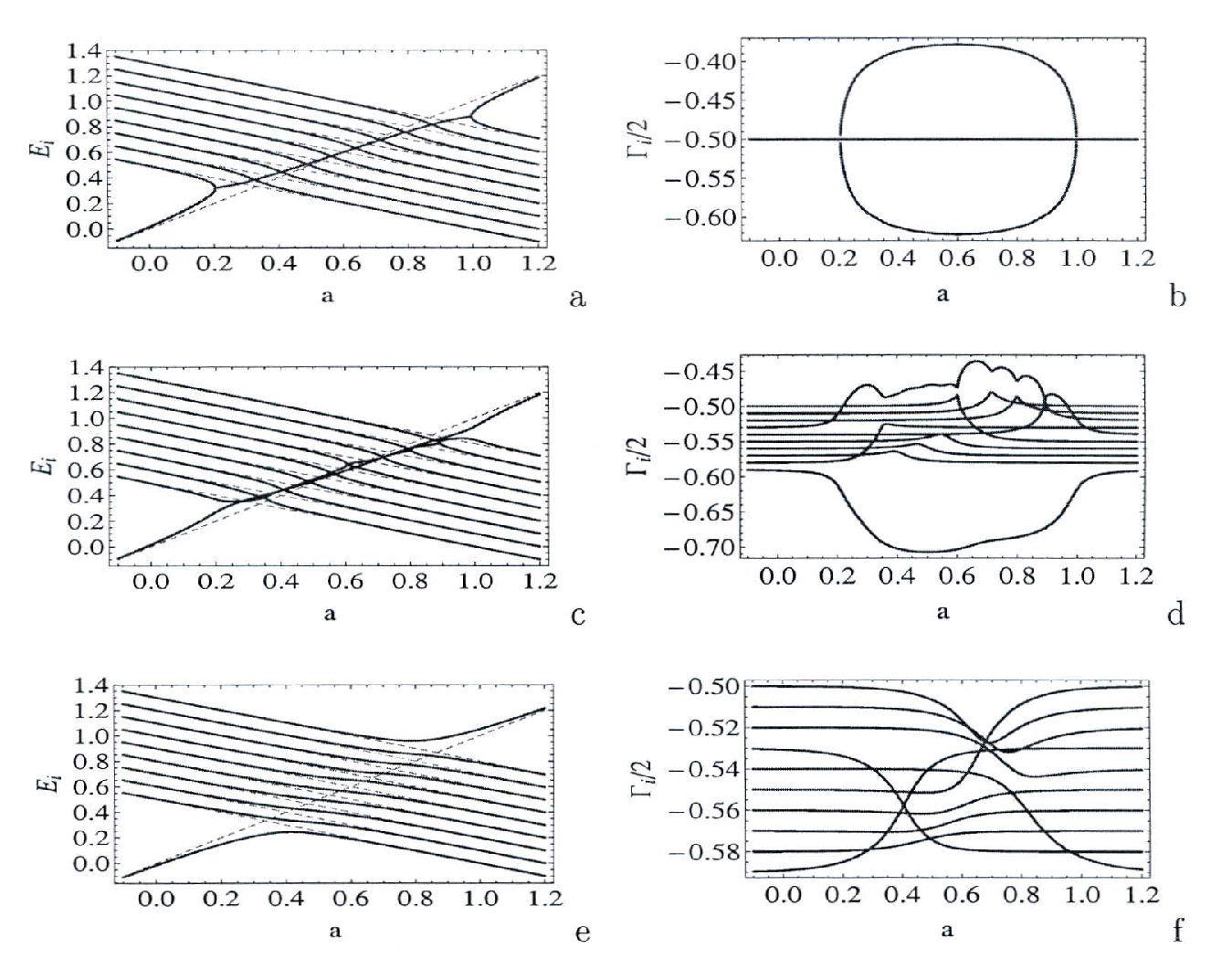}
\end{center}
\caption{\footnotesize
Energies $E_i$ and widths $\Gamma_i/2$ (full lines) 
of $N=10$ states coupled to $K=1$ channel.
The parameters of the subfigures are 
$\gamma_i/2 = 0.5$ (a,b); ~$\gamma_i/2 = 0.50; ~0.51; ~0.52; ~0.53;
~0.54; ~0.55; ~0.56; ~0.57; ~0.58; ~0.59$ (c to f);
~~$\omega=0.07~i$ (a to d); ~0.07 (e,f).
Further parameters:
~$e_i=1-a/2; ~1.1-a/2;
~1.2-a/2; ~1.3-a/2; ~0.9-a/2; ~0.8-a/2; ~0.7-a/2; ~0.6-a/2;
~0.5-a/2; ~a$. 
~The dashed lines show $e_i(a)$. 
}
\label{fig9}
\end{figure}

\begin{figure}[ht]
\begin{center}
%\includegraphics[width=6cm,height=3cm]{Multilevel-13a-gen.eps} a
%~~\includegraphics[width=6cm,height=3cm]{Multilevel-13b-gen.eps} b
%\\[.5cm] 
%\includegraphics[width=6cm,height=3cm]{Multilevel-55a-gen.eps} c
%~~\includegraphics[width=6cm,height=3cm]{Multilevel-55b-gen.eps} d
%\\[.5cm]
%\includegraphics[width=6cm,height=3cm]{Multilevel-66a-gen.eps} e
%~~\includegraphics[width=6cm,height=3cm]{Multilevel-66b-gen.eps} f
%\\[.5cm]
%\includegraphics[width=6cm,height=3cm]{Multilevel-67a-gen.eps} g
%~~\includegraphics[width=6cm,height=3cm]{Multilevel-67b-gen.eps} h
\includegraphics[width=14cm,height=14cm]{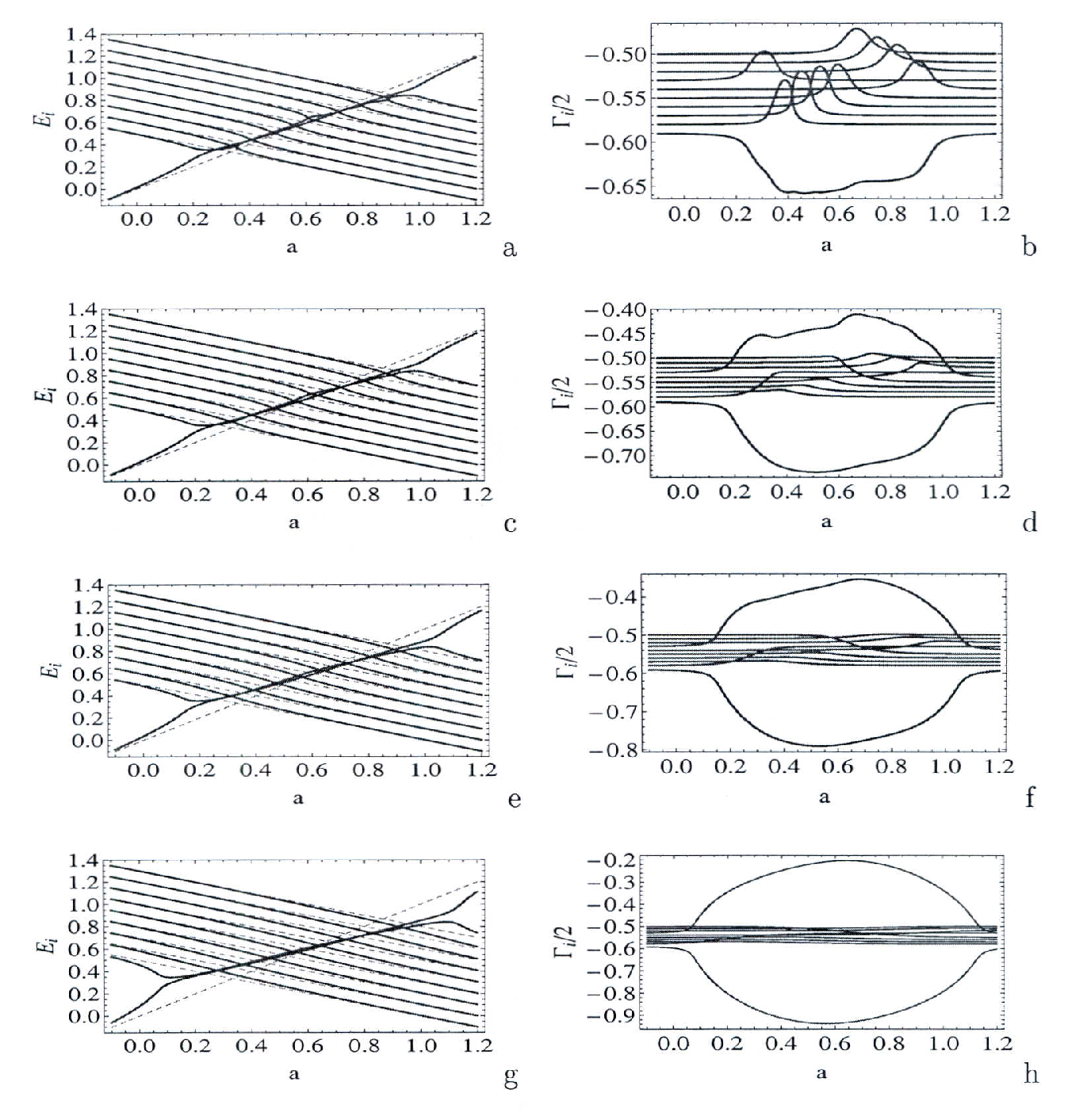}
\end{center}
\caption{\footnotesize
The same as Fig. \ref{fig9} but   ~$\omega = 0.05~i$ (a,b); 
~$\omega = 0.08~i$ (c,d); ~$\omega = 0.10~i$ (e,f);  ~$\omega =
0.15~i$ (g,h).
}
\label{fig10}
\end{figure}

We continue our studies with the matrix (\ref{int2}) by choosing 
$N=10$ states coupled to $1$ channel. The eigenvalue trajectories
$E_i(a)$ and $\Gamma_i(a)$
are shown in Figs. \ref{fig9} and \ref{fig10}. In these cases, one of
the states crosses in energy the remaining nine states one by one.
The nine states have the same parametric energy dependence of $a$ and
are shifted equidistantly relative to one another.

\begin{figure}[ht]
\begin{center}
%\includegraphics[width=6cm,height=3cm]{asymetric-1a.eps} a
%~~\includegraphics[width=6cm,height=3cm]{asymetric-1b.eps} b
%\\[.5cm] 
%\includegraphics[width=6cm,height=3cm]{asymetric-2a.eps} c
%~~\includegraphics[width=6cm,height=3cm]{asymetric-2b.eps} d
%\\[.5cm]
%\includegraphics[width=6cm,height=3cm]{asymetric-3a.eps} e
%~~\includegraphics[width=6cm,height=3cm]{asymetric-3b.eps} f
\includegraphics[width=14cm,height=11cm]{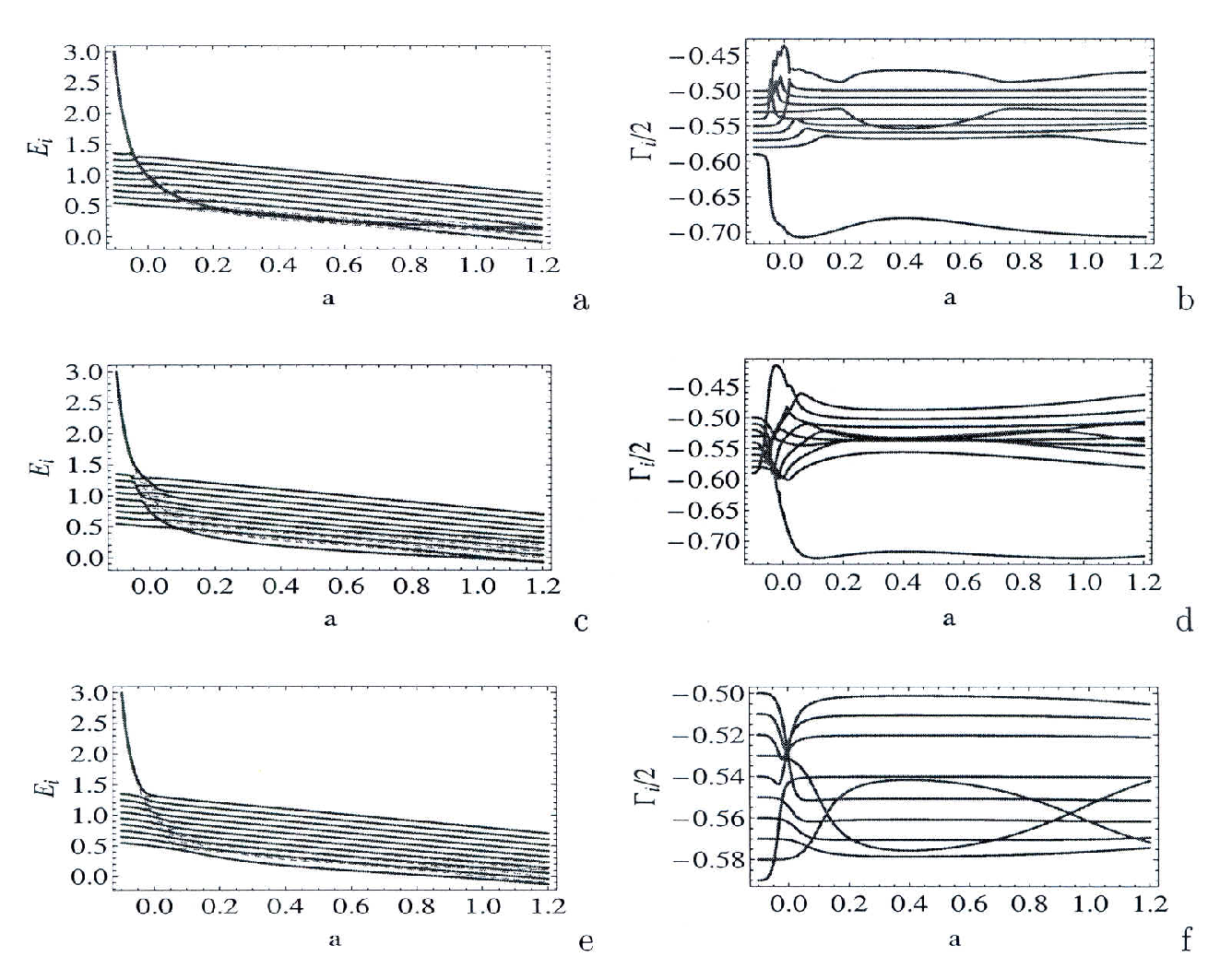}
\end{center}
\caption{\footnotesize 
The same as Fig. \ref{fig9} but $e_{10}= 0.15/(0.15+a)$; 
~~$\omega = 0.07~i$ (a, b); 
~$\omega =(1+i)~0.07$ (c, d) and ~$\omega=0.07$ (e,f). 
}
\label{fig11}
\end{figure}

In Fig. \ref{fig9}.a, b we show the results obtained with $\omega =
0.07~i$ and  $\gamma_i = 0.5$ for  all states. The results are similar to
those for the 4-level case (Figs. \ref{fig3}.c, d).    
Two exceptional points appear as well as width bifurcation\,: 
the width of one of the states is much larger while that of another one
is much smaller  than
the widths of all the other states in the whole critical parameter
range between the two exceptional points. 

In Figs. \ref{fig9}.c to f, the widths $\gamma_i$ differ from one another.
We compare the results obtained with an imaginary
and with a real coupling constant, $\omega = 0.07 ~i$ and $0.07$,
respectively. When $\omega$ is real (Figs. \ref{fig9}.e, f), the
energy  trajectories $E_i(a)$ are very regular\,:
avoided level crossings appear one by one with all nine states.
An exceptional point can be found only by analytical continuation into
the continuum (see \cite{top}). 
The width trajectories $\Gamma_i(a)$ are less 
regular\,: they cross at several parameter values $a$.
According to the results obtained for a smaller number of states in
sections \ref{2} to \ref{f4}, the widths do not bifurcate 
when $\omega$ is real. 

When $\omega$ is imaginary, the energy trajectories  $E_i(a)$ are also regular
(Fig. \ref{fig9}.c). Fig. \ref{fig9}.d
shows nicely how widths continue to bifurcate such that  the broad
state remains broad and the narrow state remains narrow for all $a$ in 
the critical region. 
Mostly, the widths of two states are different from one another  
when the energy trajectories  cross. The states can, nevertheless, be
exchanged as can be seen from Fig. \ref{fig9}.c.  

In Fig. \ref{fig10}, we show results obtained with the
imaginary coupling constant $\omega = 0.05 ~i$ 
(being smaller than in Figs. \ref{fig9}.c, d) 
as well as those obtained with the larger values
$\omega = 0.08  ~i, ~0.10 ~i, ~0.15 ~i$ (Figs. \ref{fig10}.a,b and 
Figs. \ref{fig10}.c to h, respectively). We
see width bifurcation similar as in Fig. \ref{fig3}.d 
for $N=4$ states and in Fig. \ref{fig9}.b for 10 states with equal
widths $\gamma_i$. In any case, width bifurcation
appears in the whole critical parameter range
between the two exceptional points. The width bifurcation occurring
in the case of 10 levels is stronger than that with a smaller number
of levels (compare section \ref{v4}). 
The 9 states crossed by the state 10 are exchanged 
in the critical parameter region as can be seen from the
corresponding Figs. \ref{fig10}.a, c, e, g. 

The results shown in Figs. \ref{fig9} and   \ref{fig10} follow from
calculations with a toy model being symmetrical in relation to the
crossing states. In the critical parameter range, the states are
exchanged in all cases but 8 of them do not contribute 
actively to the width bifurcation
when $\omega$ is imaginary. This result follows from the 
high symmetry of the exceptional points in relation to the crossing states,
i.e. from the linear
dependence of the energy of the crossing state on the parameter 
$a$, ~$e_{10}(a)=a$. The 8 states play the role of 'observers' similarly as
discussed in a
toy model with 4 states \cite{fdp2} and in a realistic model with 3
states \cite{rosad}.

In order to exclude the high symmetry of the exceptional points in
relation to nearby states we show in Fig. \ref{fig11} the results
of another version of the
toy model. Instead of  $e_{10}(a)=a$ in Figs. \ref{fig9} and
\ref{fig10}, we use $e_{10}= 0.15/(0.15+a)$ in Fig. \ref{fig11}. 
The width $\Gamma_i$ of one of the states becomes large in the
critical parameter range when $\omega$ is imaginary or complex
(Figs. \ref{fig11}.b and d, respectively), while the widths of all the
other states are much smaller. This result reflects the situation
observed in many realistic cases (see e.g. the review \cite{top}
and also \cite{hierarch}). A
separation of a short-lived state from the other ones does not appear
when $\omega$ is real (Fig. \ref{fig11}.f).

\section{Discussion of the results}
\label{disc}

In sections \ref{2} to \ref{v10}, we showed results obtained numerically
for the eigenvalues ${\cal E}_i= E_i-i/2~\Gamma_i$ and eigenfunctions 
$\Phi_i$ by using the matrix (\ref{form1}) (or (\ref{int2}))
with 2, 4 and 10 levels, respectively. Only the energies
$e_i$ of the states are varied as a function of a certain parameter $a$.
The widths $\gamma_i$ are assumed to be constant,  the interaction
$\omega_{ij}$ between the levels $i$ and $j$ is either zero or
simulated by the Gaussian
distribution (\ref{int7}) where $\omega$ is the same for all
states (and also independent of $a$), 
and the self-energy terms are considered to be included into the 
diagonal matrix elements. Most interesting is the one-channel case
for different $N$, e.g. the matrix (\ref{int2}) for  $N=4$.

This toy model describes well the generic features of open quantum
systems. The  $\omega_{ij}$ are complex. They stand for the
interaction of the states
via the environment (continuum of scattering wavefunctions)\,:
Re($\omega_{ij}$) arises from the principal value integral (\ref{form11})
and Im($\omega_{ij}$) is the residuum (\ref{form12}).
Both parts play an important role for the dynamics of the open quantum
system at high level density where the resonance states overlap. The 
influence of Re($\omega_{ij}$) onto eigenvalues  ${\cal E}_i$ and eigenfunctions 
$\Phi_i$ is studied numerically exact for $N=2$ levels in an earlier paper
\cite{ro01}. It  causes the  avoided crossing of the states
which is well known for discrete as well as for narrow resonance
states. 

Equations (\ref{int6a}) to  (\ref{int6d}) show analytically the main 
difference between Re($\omega_{ij}$) and Im($\omega_{ij}$) in the
2-level case when the widths of the two states are equal,
$\gamma_1=\gamma_2$. According to (\ref{int6a}),
an exceptional point can be found  only by
analytical continuation into the continuum 
when  $\omega_{ij}$ is real. Thus, the two states avoid crossing as it is very
well known for discrete as well as for narrow resonance states \cite{ro01}. 
In contrast, Im($\omega_{ij}$) causes two exceptional points according
to  equation (\ref{int6b}). Most interesting is the width bifurcation
arising in between the two exceptional points 
according to  (\ref{int6d}). 

The analytical results   (\ref{int6a}) to  (\ref{int6d}) are well
reproduced in our numerical calculations for the eigenvalues ${\cal E}_i$, 
Figs. \ref{fig3}.a, b. 
Also the eigenfunctions $\Phi_i$ show the two exceptional
points\,: $|b_{ij}|^2 \to \infty$ in approaching them,  Fig. \ref{fig6}.a. 
In between the two exceptional points, the wavefunctions of the two
states are strongly mixed. The mixing is 1:1 in the middle between the 
two singular
points. Beyond the critical region between the two exceptional points, 
the $|b_{ij}|^2$ approach 1 when $i=j$ and 0 when $i\ne j$. Here, 
the two states may be exchanged, at most.
The figures 3.a,b and 6.a represent the basic process of width
bifurcation according to the analytical expressions 
(\ref{int6c}) and (\ref{int6d}).

In sections \ref{v4}, \ref{f4} and \ref{v10} of the present paper, 
numerical results for 4 and 10 states are shown under similar conditions, 
i.e. with equal or nearly equal $\gamma_i$  for all states\,:
Figs. \ref{fig3}.c to f and Figs. \ref{fig6}.b and c for 
the eigenvalues and eigenfunctions, respectively, of 4 states, and
Figs. \ref{fig9}.a, b for the eigenvalues of 10 states. The
development of width bifurcation  as a function of
increasing imaginary coupling vector $\omega$ can be seen 
in Fig. \ref{fig10} for $N=10$ states when
the single widths $\gamma_i$ are different from one another. The
uniform width bifurcation in Fig. \ref{fig9}.b appears at a smaller value
of Im$(\omega)$ than in Fig. \ref{fig10}.h due to the different values
of the single $\gamma_i$ in the last case. Generally, the width
bifurcation is stronger when the number of states is larger.

The influence of  Re($\omega_{ij}$) onto the eigenvalue trajectories and
eigenfunctions is also shown for $N=4$ and 10
(Figs. \ref{fig4}.c, d for the eigenvalues
of 4 states,  Fig. \ref{fig7}.b for the
eigenfunctions of 4 states  and Figs. \ref{fig9}.e, f for the eigenvalues
of 10 states). Under the influence of Re($\omega_{ij}$), the states 
avoid crossing in energy and the widths do not bifurcate. When the 
calculation is performed with complex  coupling vector
(Re$(\omega_{ij}\ne 0)$ {\it and}  Im$(\omega_{ij}\ne 0)$), 
the widths bifurcate with some shift of the position of the maximum
relative to that of the minimum (Figs. \ref{fig4}.a, b)
and a larger parameter range of mixed wavefunctions (Fig. \ref{fig7}.a). 

According to the results presented in sections \ref{v4} to \ref{v10} 
the eigenvalue trajectories are strongly influenced by the values
of the external (fixed) parameters when the
resonances overlap and exceptional points determine the dynamics of
the system.  However, the eigenvalue trajectories are almost
independent of one another when the degree of resonance overlapping is
small. Examples are Figs. \ref{fig3}.g to j and
\ref{fig4}.e to h for the eigenvalues of, respectively, two and
four resonance states and the corresponding Figs. \ref{fig6}.d,e 
and  \ref{fig7}.c,d for the eigenfunctions of these states. 
These results show the strong influence of external parameters
onto the dynamics of the system at high level density
which is known from the study of
realistic cases. An example is the enhanced transmission through
microwave cavities when the formation of whispering gallery modes 
is supported by the manner the leads are attached to the cavity \cite{whisp}. 

Width bifurcation is directly related to the {\it alignment} of one of the
$N$ resonance states of the open quantum system
to a decay channel ($K=1$) with the consequence that it
becomes short-lived while  other states become {\it trapped}
(long-lived), i.e. more or less decoupled from the environment. 
Mathematically, the alignment of a resonance state
is possible since the eigenfunctions of a non-Hermitian operator are
biorthogonal and their phases (relative to those of the eigenfunctions
of the other states) are not rigid in approaching an exceptional
point, see equation (\ref{eif11}).
Width bifurcation appears therefore in our calculations when 
$K=1$ and $N=4$ or 10. This
corresponds to realistic situations in which usually $K<N$. 
Width bifurcation  does, however, not appear when $N=4$ states are
coupled to $K=4$ channels (Fig. \ref{fig5}) although
the wavefunctions are strongly mixed in the critical parameter region
also in this case (Fig. \ref{fig8}).  

An artifact of our model calculations is the assumption of 
equidistant energies $e_i$
of all but one state which are crossed by one state the energy 
of which depends linearly on the parameter $a$. Thus,   
the critical points are symmetrically in relation to the two
neighboring states, and $N-2$ states do
not contribute actively to the width bifurcation taking place in the
whole system. In Fig. \ref{fig11}, we show results obtained when the symmetry
is somewhat disturbed (as it is usually the case in realistic
systems). In such a case, the width of one state
separates from those of {\it all} the other states. That means, all
nine states remain trapped (decoupled) in the whole critical parameter range.  
This result corresponds to results obtained theoretically as well as
experimentally in different realistic systems (see section 4 
in the review \cite{top}).

It should be added here, that asymptotically (beyond the critical
parameter region) the states can be exchanged at most. This is the
case, indeed, in all our calculations. All redistribution processes 
caused by the exceptional points take place only in the critical parameter 
region. This holds true for real as well as for imaginary coupling coefficients
$\omega_{ij}$ and can be seen also in the mixing coefficients $b_{ij}$
which approach 1 or 0 when $i=j$ and $i\ne j$, respectively,
beyond the critical parameter region. 
The two cases with real and imaginary $\omega$
differ, however, by the length of the critical region.

\section{Conclusions}
\label{concl}

The results of the present paper show that exceptional points cause
width bifurcation because the coupling of the states via the
environment  (continuum of scattering wavefunctions) is complex. 
An example are the bound states in the continuum appearing at
 a {\it finite} value of the coupling strength between system and
 environment when the interplay between internal and external
 interaction is taken into account, see section 4.4 of the review \cite{top}.
Of particular interest is the width bifurcation appearing at high level
density. Here, many exceptional points
are near to one another. As shown in the present paper, width bifurcation 
causes, under this condition, a splitting of the system into two
parts one of which exists in the short-time scale while the other one
appears in the long-time scale. The two parts can not be observed
together. The long-lived states occur as fluctuations in the
short-time scale while the short-lived states perform a smooth
background in the long-time scale. 

This phenomenon is known in literature from theoretical as well as
experimental studies on different small quantum systems 
and systems equivalent to them (see examples in the review \cite{top} and 
in \cite{fdp2,pragueconf,jmp}). In many-body open quantum systems,
it is called mostly {\it dynamical phase transition}. It is known also
in optics where it is called {\it superradiant phase transition}
according to Dicke \cite{dicke}. In PT-symmetric systems, the phase
transition is called mostly {\it PT-symmetry breaking} \cite{ptsymm}.  

Common to all these studies is that the dynamical phase transition is very
robust when the necessary conditions are fulfilled, i.e. when the
level density is high and the
dynamics of the system is determined by exceptional
points. Although the existence 
of exceptional points is decisive (as shown in the present 
paper), the dynamical phase transition does {\it not} appear at the 
parameter value which gives the 
position of the exceptional point itself. Instead, the phase transition
occurs in the neighborhood of one (or several) exceptional points.
Such a result is known from different experimental studies on concrete 
realistic systems. It is discussed in e.g. \cite{jopt}. The  
results of the present paper show  that the critical parameter
range is determined by the distance between (at least) two exceptional points.

The dynamical phase transition appearing in small quantum systems at
high level density, causes a new understanding of time as discussed in
\cite{fdp1}. Time which is characteristic of the system, 
is inverse proportional to the widths $\Gamma_i$ and 
related therefore directly to the non-Hermitian part of the Hamilton
operator. The widths do not increase limitless as shown  in the
present paper. Instead, the system is
dynamically stabilized\,: the lifetimes of the states of the system are  
increased and the system as a whole is stabilized
by ejecting the  short-lived state from the system at 
the dynamical phase transition. Thus, the time characteristic of the
system is bounded from below in a similar manner as the energy. 
Beyond the dynamical phase transition, all 
states have lost their original spectroscopic features and the number
of states is reduced. 

This fact is very well known in nuclear physics
(although very seldom interpreted in this manner). The properties of
compound nuclei are described well by the {\it Unified theory of
  nuclear reactions} developed by Feshbach \cite{feshbach}, which
contains both the short-lived direct reaction part  and the
long-lived compound nuclear reaction part. However, Feshbach introduces
statistical assumptions in order to describe the long-lived compound 
nuclear states. This is in contrast to the calculations in the present 
paper where the long-lived states
are described  {\it without} any statistical assumptions.
Instead the long-lived states are shown by us to arise from 
width bifurcation causing a dynamical stabilization of the system at
high level density, a phenomenon that is caused by exceptional points.

Many of these results seem to be counterintuitive. They allow us, however, 
to explain some unexpected experimental results. For example, the 
crossover from the mesoscopic to a universal phase for transmission in
quantum dots observed experimentally \cite{laps1}  can be explained
qualitatively by the formation of the short-lived resonance state at the
dynamical phase transition. The results of the present paper support
this interpretation  \cite{laps3} of the experimental results. 
Another example is the experimental observation \cite{kanter} of non-statistical
effects in nuclear reactions on middle-heavy nuclei. The data show
directly the formation of different time scales in the system at high
level density.  Other examples are discussed in \cite{top,fdp1,jopt}. 

The long-lived states beyond the dynamical phase transition are
strongly mixed and show  chaotic features. 
The random matrix theory is surely applicable only
after averaging over different decay channels (as usually done),
i.e. it is not applicable to the description of small systems coupled
to one channel (and, respectively, to two channels in the case of 
transmission through the system). 
A theoretical analysis of the spectra (without any statistical
assumptions)   by restricting to the results obtained from
one decay channel according to the recent experimental data 
on compound nuclei \cite{koehler} is not performed up to now. 
On the basis of the results of the present paper,
it can be stated today only the following. The mixing 
of the long-lived states (beyond the dynamical phase transition) is caused
by complex many-body forces via the continuum of scattering wavefunctions
(simulated by the $\omega_{ij}$ in our calculations) and not by two-body
forces. This may provide an explanation of the fact why  compound
nucleus spectra (after averaging over different channels and beyond 
decay thresholds) may be described  by a Gaussian orthogonal ensemble
(containing many-body forces), but not by a two-body random
ensemble. This point should be studied in future in more detail. 

In order to prove the role of exceptional points in an open
many-level quantum system it is highly interesting to study 
experimentally time symmetry breaking caused by the influence 
of a nearby state onto an exceptional point.
Theoretical studies with a symmetrical influence (as in most
calculations of the present paper)  can surely not describe the
properties of realistic systems with broken time symmetry. Symmetry breaking 
influences not only width bifurcation (as shown in Fig. \ref{fig11}) 
but will prove, above all, the irreversible processes caused by 
exceptional points in open quantum systems. These processes are 
decisive, among others, for the dynamical stabilization of quantum
systems and the formation of  quantum chaos.

Finally we remark that the sensitive dependence of the dynamics of an
open quantum system on the value of external parameters
(as shown in the present paper) can be used in
order to construct small systems with desired properties. This point
is important for applications.

\vspace{1cm}

$^*$ email: eleuchh@iro.umontreal.ca\\
\hspace*{.6cm}$^{**}$ email: rotter@pks.mpg.de \\


\begin{thebibliography}{99}
\bibitem{comment} In studies by other researchers, the factor $i$ in 
(\ref{eif5}) does not appear. This difference is
discussed in detail and compared with experimental data in the
Appendix of \cite{fdp1} and in section 2.5 of \cite{top}.
\end{thebibliography}
\end{document}